%% file: coherence_time_and_beamwidth_v4.tex
\documentclass[journal,onecolumn,draftcls,12pt]{IEEEtran}
\usepackage{fixltx2e} 

\usepackage{color}
\usepackage[usenames,dvipsnames]{xcolor}
\usepackage{cite}
\usepackage{subfigure}
\usepackage[cmex10]{amsmath}
\usepackage{amsfonts}

\input{new_command}

\newcommand{\Tc}{{T_{\mathrm{c}}}}
\newcommand{\TB}{{T_{\mathrm{B}}}}
\newcommand{\fD}{{f_{\mathrm{D}}}}
\newcommand{\fDsq}{{f_{\mathrm{D}}^2}}

\newcommand{\kr}{{k_{\mathrm{r}}}}
\newcommand{\krsq}{{k_{\mathrm{r}}^2}}

\newcommand{\rmlos}{{\mathrm{LOS}}}
\newcommand{\rmnlos}{{\mathrm{NLOS}}}

\newcommand{\PL}{{\mathsf{PL}}}
\newcommand{\fc}{{f_{\mathrm{c}}}}
\newcommand{\rms}{{\mathrm{s}}}

\newcommand{\redtext}[1]{{#1}}
\newcommand{\bluetext}[1]{{#1}}
\newcommand{\greentext}[1]{{#1}}

\begin{document}
%
\title{The Impact of Beamwidth on Temporal Channel Variation in Vehicular Channels and its Implications}
%
%
%

\author{Vutha Va, Junil Choi, 
       and~Robert W. Heath Jr. 
\thanks{The authors are with the Wireless Networking and Communications Group, the University of Texas at Austin, TX 78712-1687 USA (e-mail: vutha.va@utexas.edu, junil.choi@utexas.edu, rheath@utexas.edu).}
\thanks{This material is based upon work supported in part by the National Science Foundation under Grant No. NSF-CCF-1319556, the U.S. Department of Transportation through the Data-Supported Transportation Operations and Planning (D-STOP) Tier 1 University Transportation Center, the Texas Department of Transportation under Project 0-6877 entitled ``Communications and Radar-Supported Transportation Operations and Planning (CAR-STOP)", and by a gift from TOYOTA InfoTechnology Center, U.S.A., Inc. Part of this work has been presented at the 2015 IEEE 82nd Vehicular Technology Conference, Sep. 2015 \cite{Va2015}. }
}

%



\maketitle

\begin{abstract}
Millimeter wave (mmWave) has great potential in realizing high data rate thanks to the \redtext{large spectral channels}. It is considered as a key technology for the fifth generation wireless networks and is already used in \redtext{wireless LAN} (e.g., IEEE 802.11ad). Using mmWave for vehicular communications, however, is often viewed with some skepticism due to a misconception that the Doppler spread would become too large at these high frequencies. This is not true when directional beam is employed for communications. \redtext{In this paper}, closed form expressions relating the channel coherence time and beamwidth \redtext{are} derived. Unlike prior work that assumed perfect beam pointing, the pointing error due to the receiver motion is incorporated to show that there exists a non-zero optimal beamwidth that maximizes the coherence time. 
To investigate the mobility effect on the beam alignment which is an important feature in mmWave systems, a novel concept of beam coherence time is defined. The beam coherence time, which is an effective measure of beam alignment frequency, is shown to be much larger than the conventional channel coherence time and thus results in reduced beam alignment overhead. 
Using the derived correlation function, the channel coherence time, and the beam coherence time, an overall performance metric considering both the channel time-variation and the beam alignment overhead is derived. Using this metric, it is shown that beam alignment in every beam coherence time performs better than the beam alignment in every channel coherence time due to the large overhead for the latter case. 
\end{abstract}

\begin{IEEEkeywords}
Channel coherence time, Vehicular communication, Doppler spread, Millimeter wave, Beam alignment, Beamforming.
\end{IEEEkeywords}

%
\IEEEpeerreviewmaketitle

\section{Introduction}

\greentext{Vehicular environments offer a fertile ground for innovative applications of wireless communications,
ranging from safety to traffic efficiency to entertainment. These new applications are pushing the boundaries of what can be done with conventional wireless technologies for vehicular applications. For example, exchanging raw sensor data or high quality multimedia could require gigabit-per-second data rates. The state-of-the-art approach for communicating in vehicular environments is dedicated short-range communication (DSRC) \cite{Kenney2011}. This standard though offers \bluetext{data rates on the order of several megabits per second \cite{Kenney2011,Jiang2008}}. Fourth generation cellular offers higher data rates, though not gigabits-per-second and also has longer latency \cite{Ghosh2010}. This motivates developing new approaches for communicating between vehicles at much higher data rates.} 

Millimeter wave (mmWave) has great potential in realizing gigabits-per-second data rates by taking advantage of the huge spectral bandwidths at these high frequencies. It is being considered as a potential candidate for the fifth generation (5G) cellular networks \cite{Andrews:What-will-5g-be:14}, and it is already in use in WPAN/WLAN standards such as WirelessHD \cite{wirelessHD-spec} and IEEE 802.11ad \cite{802.11ad}. 
One main concern in applying mmWave to vehicular environments is the severity of the Doppler effect due to the small wavelengths at mmWave frequencies. Based on the \redtext{Clarke-Jakes} power angular spectrum (PAS), it \redtext{follows} that the channel coherence time $\Tc$ is inversely proportional to the maximum Doppler frequency $\fD$, i.e., $\Tc\simeq \frac{1}{\fD}$ \cite{Goldsmith2005}. This \redtext{implies} that by moving from a typical cellular frequency at around 2 GHz to a mmWave frequency at 60 GHz, one would expect a 30x decrease in the channel coherence time, which would greatly challenge the PHY layer design. 
\redtext{As argued in this paper, this} is in fact inaccurate for mmWave systems that use directional antennas (or beams) creating angular selectivity in the incoming signal. 

Directional reception can increase the channel coherence time \cite{Durgin2000,Chizhik2004}. The Clarke PAS assumes that the incoming signals arrive uniformly over all the 360$^\circ$ angular range, which holds under rich scattering environments with omni-directional reception. To compensate for the increased path loss due to the shrinking antenna size at the mmWave frequencies, beamforming is widely accepted as a necessary component in enabling mmWave communication systems \cite{mmWavebook}. With directional reception, the incoming signals are limited to a given range of angles. Each angle can be mapped to a Doppler frequency shift, and thus this also means that the Doppler frequency shifts are limited to a certain frequency range with directional reception. Since the average frequency shift can be corrected using standard frequency offset correction methods, this leads to reduced Doppler spread and thus an increased coherence time. This property has been exploited in \cite{Chizhik2004} and \cite{Norklit1999} to mitigate the Doppler spread.



Using directional transmission and reception can help slow down the channel variation at the expense of beam alignment overhead, i.e., \redtext{loss in system spectral efficiency due to radio resources consumed to find} the best transmit and receive directions. The channel is approximately constant during a channel coherence time. If beam alignment is done in every channel coherence time, it is assured that the best transmit and receive beams are always chosen. This approach, however, will cause excessive beam alignment overhead if the channel coherence time is not long enough \greentext{to take advantage of the fully aligned beams.} The physical beam can be associated with a propagation path (similar to a path of a ray in the ray-tracing model in \cite{Tse2005}) whose angle of arrival could change much slower than the fading channel coefficient. We define a beam coherence time to capture this effect. One natural question is how much is the loss if beams are realigned at this slower speed? 
\greentext{We call the beam realignment in every channel coherence time the short-term realignment, and the beam realignment in every beam coherence time the long-term realignment.}
\greentext{This naming roots from our numerical results which show that the beam coherence time can be an order of magnitude longer than the channel coherence time.} We will show in Section \ref{sec:realignment_duration} that the overhead of the short-term realignment costs more than the gain, and the long-term realignment actually performs better. 



\redtext{The main objective of this paper} is to understand the potential of the mmWave vehicular communications using directional beams in fast changing vehicular environments. Although our channel model does not specify the carrier frequency, our focus is on mmWave bands and accordingly all of our numerical examples use parameters from the 60 GHz band. Our contributions are \redtext{summarized} as follows.
\begin{itemize}
\item We derive the channel \redtext{temporal} correlation function taking into consideration both the pointing error due to the receiver motion and Doppler effect. Based on the obtained correlation function, we derive the channel coherence time and show its connection to the receive beamwidth and the pointing direction. Our results show that there exists a non-zero optimal beamwidth that maximizes the channel coherence time unlike prior work that assumes perfect beam pointing.
\item We \redtext{propose} a novel definition of beam coherence time that is used as the basis for studying the long-term beam realignment. This enables us to study the required beam alignment overhead and allows the comparison between the short- and long-term beam realignment.  
\item We investigate the choice of the beam realignment duration taking both the beam alignment overhead and the loss due to the channel time-variation into consideration. We show that long-term beam realignment performs better and thus the beams can be realigned every beam coherence time, not every channel coherence time. 
\end{itemize}
\greentext{Our prior work in \cite{Va2015} covered part of the first contribution. In \cite{Va2015}, we derived the channel temporal correlation function and the channel coherence time for the non-line-of-sight (NLOS) channels only, while in this paper we also consider the line-of-sight (LOS) channels. Furthermore, we propose a definition of beam coherence time and investigate its implication as described in the second and third bullet above.}


Relevant \redtext{prior} work includes \cite{Teal2002,Rad:Impact-non-isotropic-scattering-directional-antenna:08,Zajic:Space-time-correlated-mobile-to-mobile:08,Xiang2009} that characterized the channel correlation under non-isotropic scattering environments. In our \greentext{paper}, the angular selectivity of the incoming signals is controlled by using narrow receive beams. Although the underlying phenomenon is different, non-isotropic scattering environments also cause angular selectivity of the incoming signals thus both have similar effect only that the selectivity resulting from non-isotropic scattering cannot be controlled. Generally, there are two directions in this line of \redtext{research}: one is to provide a generalized framework that can be used for any scattering distribution \cite{Teal2002,Rad:Impact-non-isotropic-scattering-directional-antenna:08} and the other is to constrain to a given distribution that allows tractable expression for further analysis \cite{Zajic:Space-time-correlated-mobile-to-mobile:08,Xiang2009}. The work in \cite{Teal2002} presented a generalized framework to compute a spatial correlation function for general 3D scattering distributions. Their result was based on the decomposition of the plane wave into infinite sum of the spherical Bessel functions and Legendre polynomials. Although general, this infinite sum convergence can be slow and it is not amendable for further analysis. A similar approach was used in \cite{Rad:Impact-non-isotropic-scattering-directional-antenna:08} to compute correlation functions in 2D while also taking the antenna patterns into account. For the 2D case, the plane wave is decomposed into an infinite sum of the Bessel functions. Similar to \cite{Teal2002}, the obtained correlation function is intractable for further analysis. The work in \cite{Zajic:Space-time-correlated-mobile-to-mobile:08,Xiang2009} instead considered only the von Mises scattering distribution and derived closed form correlation functions using two-ring models. Our work follows this later path and adopt the von Mises distribution to represent the effective PAS. Different from \cite{Zajic:Space-time-correlated-mobile-to-mobile:08,Xiang2009}, we also incorporate the pointing error due to the receiver motion into our correlation function, which is an essential characteristic when using directional beams for vehicular environments.  

\redtext{Other prior work related to our research appears in} \cite{Durgin2000,Norklit1999,Chizhik2004,Hur2013}. The relationship between the channel coherence time and beamwidth was also studied in \cite{Durgin2000,Chizhik2004}. A general framework to compute the coherence time was derived in \cite{Durgin2000} for any PAS. The correlation was defined using the channel amplitude and the main assumption was that the channel coefficient is Rayleigh distributed. Our work defines correlation using the complex channel coefficient which takes not only the amplitude but also the phase into consideration. The work in \cite{Chizhik2004} related the coherence time with the number of the antennas assuming a linear array. A simple expression was derived for a special case when the pointing angle is $90^\circ$. \redtext{Although not explicitly relating channel coherence time and beamwidth, \cite{Norklit1999} proposed a beam partitioning for a mobile in a rich scattering environment such that each beam experiences the same amount of Doppler spread. For each beam, the mean Doppler shift is compensated and then synthesized back for further processing. This has the effect of slowing down the fading rate. } 
Note that in \cite{Durgin2000,Norklit1999,Chizhik2004}, no pointing error was considered and their results suggest that the coherence time goes to infinity when the beamwidth approaches zero. Our work incorporates pointing error due to the receiver motion and this enables us to show that there exists a non-zero optimal beamwidth that maximizes the channel coherence time. To the best of our knowledge, this paper is the first to incorporate both the Doppler and the pointing error to derive the channel coherence time. Recently, \cite{Hur2013} quantified the channel coherence time considering pointing error due to wind-induced vibration for mmWave wireless backhaul application. This definition of coherence time is similar to our beam coherence time in Section \ref{sec:beam_coherence_time}. Note that \cite{Hur2013} does not consider mobility and the source of pointing error is different from ours.

The rest of the paper is organized as follows. Section \ref{sec:model} describes our models and assumptions. Using the models, novel channel \redtext{temporal} correlation functions taking the pointing error into account are derived for both the LOS and NLOS cases in Section \ref{sec:corr_func}. Section \ref{sec:chan_coherence_time} derives the channel coherence time from the obtained correlation functions. In Section \ref{sec:beam_coherence_time}, a novel beam coherence time, which is tailored to the beam alignment concept, is defined. Based on these results, Section \ref{sec:implications} investigates some implications on the beam alignment duration.  
Finally, Section \ref{sec:conclusion} concludes the paper.


\section{Model and Assumption} \label{sec:model}
This section describes our models and assumptions. We start with the channel model and then introduce our model to incorporate the pointing error due to the receiver motion. Finally, we describe a spatial lobe model that \redtext{provides a statistical description of the angular spread of the PAS. The spatial lobe model will be used in the derivation of the beam coherence time.}%
\subsection{Channel Model}
This subsection first describes the NLOS channel, after which it will be incorporated into the LOS channel model. 
We assume a narrowband wide sense stationary and uncorrelated scattering (WSSUS) model for the NLOS channel, where the channel coefficient can be written as \cite{Goldsmith2005}
\begin{align}
\label{eq:channel_model_NLOS}
h_{\mathrm{NLOS}}(t) = \int_{-\pi}^{\pi} \sqrt{\mcP'(\alpha)G(\alpha|\mu_\rmr)}e^{\jj [\phi_0(\alpha) +\phi(\alpha) + 2\pi \fD t\cos(\alpha)] }  \rmd \alpha,
\end{align}
where $\mcP'(\alpha)$ is the PAS, $G(\alpha|\mu_\rmr)$ is the antenna pattern with the main lobe pointing at $\mu_\rmr$, $\phi_0(\alpha)$ is the phase due to the distance traveled up to time $0$, $\phi(\alpha)$ is the random phase associated with the path with angle of arrival $\alpha$, and $\fD$ is the maximum Doppler frequency. Note that all angles including $\alpha$ are defined in reference to the direction of travel of the receiver (Fig. \ref{fig:displacement_and_pointing_angle}). 
Under the uncorrelated scattering assumption, $\phi(\alpha)$ are uncorrelated and uniformly distributed in $[0,2\pi)$. For the time scale considered, it is assumed that the scatterers are stationary. This is based on the wide sense stationary assumption which is reasonable for a short period of time. 

We define the effective PAS $\mcP(\alpha|\mu_\rmr)$ as the power observed through the lens of the receive beam pattern, i.e., $\mcP(\alpha|\mu_\rmr)=\mcP'(\alpha)G(\alpha|\mu_\rmr)$. 
We assume the effective PAS can be represented by the von Mises distribution function with mean $\mu_\rmr$ given by 
\begin{align}
\label{eq:von_Mises_pdf}
\mcP(\alpha|\mu_\rmr) = \frac{1}{2\pi I_0(\kr)}e^{\kr\cos(\alpha-\mu_\rmr)},
\end{align}
where $I_0(\cdot)$ is the zeroth order modified Bessel function of the first kind, $\kr$ is the shape parameter. 
The von Mises distribution can be thought of as a circular version of the Gaussian distribution. For large $\kr$, it can be approximated by a Gaussian distribution with the same mean $\mu_\rmr$ and variance of $1/\kr$. We define the beamwidth $\theta$ by $\kr\simeq 1/\theta^2$. Our choice of the von Mises distribution function is based on two reasons: (i) its good resemblance to a real antenna pattern and (ii)  its tractability in our analysis. Some examples of the use of Gaussian probability distribution function (PDF), which is well approximated by a von Mises PDF for large $\kr$, in this context are its adoption as an antenna pattern in a 5G channel model \cite[Section 5.3.7.2]{Maltsev2014}, and its application as the model for the angle of arrival in another 5G channel model \cite{NurmelaVuokko2015}. 

Next we describe our LOS channel model. Here we introduce the LOS component \redtext{with} the channel coefficient now \redtext{taking the form}
\begin{align}
\label{eq:channel_model}
h(t) = \sqrt{\frac{K}{K+1}} h_{\mathrm{LOS}}(t) + \sqrt{\frac{1}{K+1}} h_{\mathrm{NLOS}}(t),
\end{align}
where $K$ is the Rician K factor, which determines the relative power between the LOS and NLOS components. The LOS component is modeled as
\begin{align}
\label{eq:channel_model_LOS}
h_{\mathrm{LOS}}(t) = \sqrt{G(\alpha| \mu_\rmr )} e^{-\jj \frac{2\pi}{\lambda}D} e^{\jj 2\pi \fD t \cos(\alpha_{\rmlos})} \delta( \alpha_{\rmlos} - \alpha )
\end{align}
where 
$D$ is the distance between the transmitter and the receiver at time $0$, $\alpha_{\rmlos}$ is the angle of arrival of the LOS path, and $\delta(\cdot)$ denotes the Dirac delta function \cite{Zajic:Space-time-correlated-mobile-to-mobile:08}.

\subsection{\greentext{Pointing Error due to} Receiver Motion} \label{sec:pointing_error_model}
This subsection explains how the effect of receiver displacement is modeled. This is based on the observation that if the receive beam is fixed and the receiver moves then beam misalignment will happen. Misalignment implies that the receiver sees the channel with a different lens than when properly aligned and thus the channel \redtext{temporal} correlation will be affected. 

\begin{figure}
\centering
\includegraphics[width=0.45\textwidth]{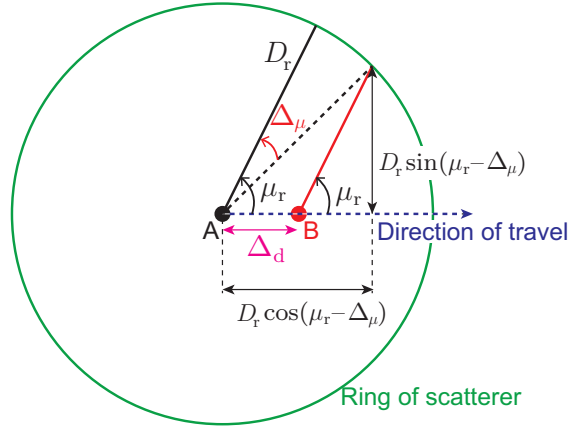}
\caption{Receiver displacement and change in pointing angle for the NLOS case. When the receiver moves from A to B, if the beam is not adaptive, then effectively the beam pointing angle changes. 
}
\label{fig:displacement_and_pointing_angle}
\end{figure}

Our model for the NLOS case follows the one-ring model, where scatterers are distributed on a ring of radius $D_\rmr$ as shown in Fig. \ref{fig:displacement_and_pointing_angle}. Assume the receiver is at point A at time $t$ and it moves with a speed $v$ along the direction of travel to arrive at point B at time $t+\tau$. The total displacement from A to B is $\Delta_\rmd(\tau)=v\tau$. 
When the receiver moves from A to B by $\Delta_\rmd(\tau)$, the receiver will see a different set of scatterers and the distances to the scatterers also change. We assume that $\Delta_\rmd(\tau) \ll D_\rmr$, \greentext{so that the displancement $\Delta_\rmd(\tau)$ has negligible effect on the path loss and capture the receiver motion effect through the pointing error $\Delta_\mu(\tau)$ as shown in Fig. \ref{fig:displacement_and_pointing_angle}.}
For notational convenience, $\Delta_\mu, \Delta_\rmd$ are used instead of $\Delta_\mu(\tau),\Delta_\rmd(\tau)$. The relationship between $\Delta_\rmd$ and $\Delta_\mu$ can be obtained from geometry as
\begin{align}
\label{eq:angle_displacement_exact}
\tan(\mu_\rmr) = \frac{D_\rmr \sin(\mu_\rmr-\Delta_\mu)}{D_\rmr \cos(\mu_\rmr-\Delta_\mu) - \Delta_\rmd}.
\end{align}
Expanding the sine and cosine term in \eqref{eq:angle_displacement_exact} and apply the small $\Delta_\mu$ approximation, i.e., $\sin\Delta_\mu\simeq\Delta_\mu$ and $\cos\Delta_\mu\simeq 1$, then \eqref{eq:angle_displacement_exact} becomes
\begin{align*}
\frac{\sin \mu_\rmr}{\cos\mu_\rmr} & \simeq \frac{D_\rmr \sin\mu_\rmr - D_\rmr \Delta_\mu \cos\mu_\rmr   }{ D_\rmr \cos\mu_\rmr + D_\rmr \Delta_\mu \sin\mu_\rmr - \Delta_\rmd } \\
 D_\rmr \Delta_\mu \sin^2 \mu_\rmr - \Delta_\rmd \sin \mu_\rmr & \simeq   - D_\rmr \Delta_\mu \cos^2\mu_\rmr.
\end{align*}
Using the identity $\sin^2 \mu_\rmr+\cos^2 \mu_\rmr=1$, a tractable approximation is obtained as
\begin{align}
\label{eq:point_ang_diff_d}
\Delta_\mu \simeq \frac{\Delta_\rmd}{D_\rmr} \sin \mu_\rmr. 
\end{align}
Since $\fD=v/\lambda$, we have $\Delta_\rmd=v\tau=\fD\lambda \tau$, where $\lambda$ is the carrier wavelength. Substituting this into \eqref{eq:point_ang_diff_d} to get
\begin{align}
\label{eq:point_ang_diff_tau}
\Delta_\mu \simeq \frac{\fD \tau}{D_{\rmr,\lambda}} \sin \mu_\rmr, 
\end{align} 
where $D_{\rmr,\lambda}=D_{\rmr}/\lambda$ is the scattering radius normalized by the carrier wavelength $\lambda$.

The same reasoning can also be applied to the LOS case by replacing the scattering radius $D_\rmr$ by the transmitter-receiver distance $D$. Let $D_\lambda=D/\lambda$ and $\alpha_{\rmlos}$ be the direction toward the transmitter (in reference to the travel direction), then
\begin{align}
\label{eq:point_ang_diff_los}
\Delta_\mu^\rmlos \simeq \frac{\fD \tau}{D_{\lambda}} \sin (\alpha_{\rmlos}). 
\end{align} 
The approximate relation \eqref{eq:point_ang_diff_tau} and \eqref{eq:point_ang_diff_los} will be incorporated with the channel model in the previous section to derive an approximate channel \redtext{temporal} correlation function in Section \ref{sec:corr_func}. 

\subsection{Channel Spatial Lobe Model} \label{sec:chan_spatial_lobe_model}
\begin{figure}
\centering
\includegraphics[width=0.4\textwidth]{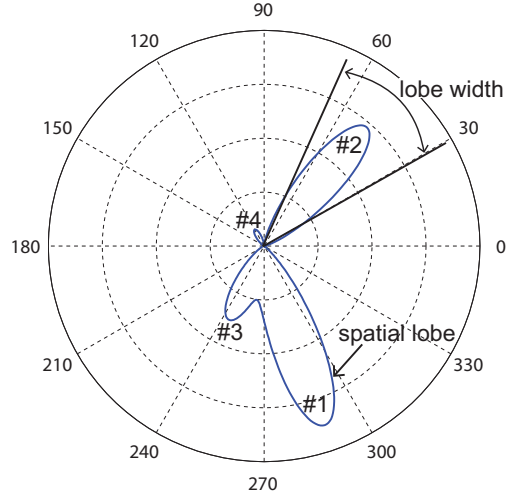}
\caption{An illustration of spatial lobes. This figure illustrates the pattern of the incoming power arriving at the receiver. The incoming power has strong spatial dependence, and it can be observed here that there are four main directions, each of which can be thought of as four clusters of scatterers. At each of these directions, there is spread forming a lobe, which is termed as a spatial lobe. In this example, lobe \#1 has the strongest power.}
\label{fig:spatial_lobe_illustration}
\end{figure}

Here we explain the spatial lobe model that will be the \redtext{basis} for our definition of the beam coherence time. Only the azimuthal plane is considered. \redtext{This model provides a statistical description of the angular spread of the PAS.} A signal transmitted from the transmitter propagates through different paths to arrive at the receiver. 
These multipaths arrive at different angles with some concentrations at certain angles which create patterns as illustrated in Fig. \ref{fig:spatial_lobe_illustration}, which are called spatial lobes. 
Four spatial lobes are shown in Fig. \ref{fig:spatial_lobe_illustration}. 
Each of them can be thought of as a result of the power coming from the multipaths corresponding to a cluster of scatterers that have similar angles of arrival. The number of spatial lobes depends on the environment and ranges from 1-6 in an urban environment measurement at 28 GHz \cite{Samimi2014}. Beam alignment is the process of finding the direction of the spatial lobe with the highest power, i.e., the lobe with the highest peak (lobe \#1 in Fig. \ref{fig:spatial_lobe_illustration}). The lobe width determines the difficulty in aligning the beam. The narrower the spatial lobe, the more difficult the alignment becomes, and the easier the beam gets misaligned due to the receiver motion. Thus this lobe width plays a fundamental role in defining the beam coherence time. 

The lobe width $\beta$ is modeled following the empirical model proposed in \cite{Samimi2014}, which uses a Gaussian distribution, i.e.,
\begin{align}
\label{eq:lobe_width_dist}
\beta \sim \Gauss(m_{\mathrm{AS}},\sigma_{\mathrm{AS}}^2).
\end{align}
The mean $m_{\mathrm{AS}}$ and the standard deviation $\sigma_{\mathrm{AS}}$ depend on the environment. The model in \cite{Samimi2014} was based on measurements in urban area, \redtext{where} $m_{\mathrm{AS}}=34.8^\circ$ and $\sigma_{\mathrm{AS}}=25.7^\circ$ were derived. 

\section{Channel \redtext{Temporal} Correlation Function} \label{sec:corr_func}
There are two possible definitions of \greentext{the} channel \redtext{temporal} correlation function. One is based on the amplitude of the channel coefficients \cite{Durgin2000} and the channel correlation function is defined by
\begin{align}
\label{eq:env_corr}
R_{|h|}(\tau)=\frac{\bbE\left[g(t)g(t+\tau)\right]-\left(\bbE[g(t)] \right)^2}{\bbE[g(t)^2] -\left(\bbE[g(t)] \right)^2},
\end{align}
where $g(t)=|h(t)|$, and $\bbE[\cdot]$ denotes the expectation operator.
The other definition is based on the complex channel coefficients themselves \cite{Chizhik2004} and is defined as 
\begin{align}
\label{eq:channel_coef_based_correlation}
R_h(\tau) = \bbE\left[h(t) h^*(t+\tau)\right],
\end{align}
where $(\cdot)^*$ denotes complex conjugate. Most modern communication systems use coherent detection, where both amplitude and phase are important. In that respect, the definition in \eqref{eq:channel_coef_based_correlation} is more natural and is the definition used in this study. It should be noted that when $h(t)$ is Rayleigh faded, the two definitions are in fact equivalent \cite[Pages 47-51]{Jakes1994}, in the sense that there is a simple relationship between the two. In particular, it can be shown that $R_{|h|}(\tau)=\frac{\pi}{4(4-\pi)}|R_h(\tau)|^2$. 

The channel model in \eqref{eq:channel_model} has both the LOS and NLOS components. 
\redtext{For the LOS component, $h_\rmlos(t)$ depends on the pointing direction, and proper normalization is needed to be consistent with \eqref{eq:channel_coef_based_correlation}. }
We still define the correlation function for the LOS component $R_{\rmlos}(\tau)$ based on the product $h_{\rmlos}(t) h_{\rmlos}^*(t+\tau)$ \redtext{but now we introduce a} normalization such that $|R_{\rmlos}(\tau=0)|=1$ and $|R_{\rmlos}(\tau\neq0)|<1$ in Section \ref{sec:corr_func_los}. Along with this definition, the correlation function of the channel is defined as
\begin{align}
\label{eq:corr_gen}
R_h(\tau) & = \frac{K}{K+1} R_{\rmlos}(\tau) + \frac{1}{K+1} R_{\rmnlos}(\tau).
\end{align}
In the followings, we 
derive the correlation function for the NLOS channel using \eqref{eq:channel_coef_based_correlation} in Section \ref{sec:corr_func_nlos} and define the correlation function for the LOS in Section \ref{sec:corr_func_los} that is consistent with the definition in \eqref{eq:channel_coef_based_correlation}. In both cases, the effect of the receiver motion is taken into account. 

\subsection{NLOS Channel Correlation Function} \label{sec:corr_func_nlos}
Here we derive the correlation function between $h_\rmnlos(t)$ and $h_\rmnlos(t+\tau)$ for the NLOS channel. 
The channel coefficients at time $t$ and $t+\tau$ are given by,
\begin{align}
\label{eq:chan_at_t}
h_\rmnlos(t) & = \int_{-\pi}^{\pi} \sqrt{\mcP(\alpha|\mu_\rmr)}e^{\jj [\phi_0(\alpha) +\phi(\alpha) + 2\pi \fD t\cos(\alpha)] }  \rmd \alpha, \\
\label{eq:chan_at_t_tau}
h_\rmnlos(t+\tau) & = \int_{-\pi}^{\pi} \sqrt{\mcP(\alpha|\mu_\rmr+\Delta_\mu)}e^{\jj [\phi_0(\alpha) +\phi(\alpha) + 2\pi \fD (t+\tau)\cos(\alpha)] }  \rmd \alpha,
\end{align}
where we have incorporated the pointing angular change due to the receiver motion in the peak direction of the effective PAS, which is now $\mu_\rmr+\Delta_\mu$ instead of $\mu_\rmr$ in \eqref{eq:chan_at_t_tau}. Plugging these into the definition in \eqref{eq:channel_coef_based_correlation},
\begin{align}
R_{\rmnlos} &(\tau)  = \bbE\left[ \int_{-\pi}^{\pi}\int_{-\pi}^{\pi}\sqrt{\mcP(\alpha_1|\mu_\rmr)\mcP(\alpha_2|\mu_\rmr+\Delta_\mu)} e^{\jj(\phi_0(\alpha_1)+\phi(\alpha_1)-\phi_0(\alpha_2)- \phi(\alpha_2)-2\pi \fD \tau \cos(\alpha_2))} \rmd \alpha_1 \rmd \alpha_2 \right] \nonumber \\
& = \int_{-\pi}^{\pi}\int_{-\pi}^{\pi} \sqrt{\mcP(\alpha_1|\mu_\rmr)\mcP(\alpha_2|\mu_\rmr+\Delta_\mu)} \bbE[e^{\jj(\phi(\alpha_1)-\phi(\alpha_2))}]e^{\jj(\phi_0(\alpha_1)-\phi_0(\alpha_2))} e^{-\jj 2\pi \fD \tau \cos(\alpha_2) } \rmd \alpha_1 \rmd \alpha_2 \nonumber \\
\label{eq:derivation_Rh_3}
& = \int_{-\pi}^{\pi} \sqrt{\mcP(\alpha|\mu_\rmr)\mcP(\alpha|\mu_\rmr+\Delta_\mu)}e^{-\jj 2\pi \fD \tau \cos(\alpha) } \rmd \alpha,
\end{align}
where \eqref{eq:derivation_Rh_3} follows from the uncorrelated scattering assumption. According to this assumption, $\bbE[e^{\jj(\phi(\alpha_1)-\phi(\alpha_2))}]=\bbE[e^{\jj\phi(\alpha_1)}]\bbE[e^{-\jj\phi(\alpha_2)}]=0$ for $\alpha_1\neq \alpha_2$ and $\bbE[e^{\jj(\phi(\alpha_1)-\phi(\alpha_2))}]=1$ for $\alpha_1=\alpha_2$, where $\phi(\alpha)$ is uniformly distributed in $[0,2\pi)$. Now substituting the von Mises PDF to get
\begin{align}
\label{eq:NLOS_corr_func_original}
R_{\rmnlos} (\tau) &= \frac{1}{2\pi I_0(\kr)} \int_{-\pi}^{\pi} \sqrt{e^{\kr \cos(\alpha-\mu_\rmr)}e^{\kr\cos(\alpha-(\mu_\rmr+\Delta_\mu))}} e^{-\jj 2\pi \fD \tau \cos(\alpha) } \rmd \alpha  \\
 &= \frac{1}{2\pi I_0(\kr)} \int_{-\pi}^{\pi} e^{\kr\cos\left(\alpha-\mu_\rmr-\frac{\Delta_\mu}{2}\right) \cos\left(\frac{\Delta_\mu}{2}\right) } e^{-\jj 2\pi \fD \tau \cos(\alpha) } \rmd \alpha \nonumber \\
 &=  \frac{1}{2\pi I_0(\kr)} \int_{-\pi}^{\pi} e^{k_\rmr'\cos\left(\alpha-\mu'\right)  } e^{-\jj 2\pi \fD \tau \cos(\alpha) } \rmd \alpha \nonumber \\
 &= \frac{1}{2\pi I_0(\kr)} \int_{-\pi}^{\pi}  e^{ x'\cos\alpha + y'\sin\alpha } \rmd \alpha \nonumber \\
 \label{eq:NLOS_corr_func_exact}
 & = \frac{I_0(\sqrt{x'^2+y'^2})}{I_0(\kr)}, 
\end{align}
where, 
\redtext{
\begin{align}
k_\rmr' & = \kr \cos\left(\frac{\Delta_\mu}{2}\right) \\
\mu' & = \mu_\rmr+\frac{\Delta_\mu}{2} \\
x' & = k_\rmr'\cos\mu'-\jj2\pi \fD \tau  \\
y' & = k_\rmr'\sin\mu'. 
\end{align}
}
and we have used the formula $\int_{-\pi}^{\pi}e^{a\cos c+b\sin c}\rmd c = 2\pi I_0(\sqrt{a^2+b^2})$ \cite[3.338-4]{Gradshteyn:Table-of-integral:07}. Despite the simple form of \eqref{eq:NLOS_corr_func_exact}, it is intractable for further analysis because the argument to the Bessel function involves the cosine of $\Delta_\mu$, which is also a function of $\tau$. 
Fortunately, a more tractable approximated form can be obtained for large $\kr$, where the von Mises PDF can be approximated by the Gaussian one with the variance of $1/\kr$ and the same mean. With this approximation, \eqref{eq:NLOS_corr_func_original} becomes 
\begin{align}
\label{eq:NLOS_corr_gaussian_approx}
R_{\rmnlos} (\tau) \simeq \frac{1}{\sqrt{2\pi/\kr}} \int_{-\pi}^{\pi} e^{-\frac{\kr}{4}\left( (\alpha-\mu_\rmr)^2+(\alpha-(\mu_\rmr + \Delta_\mu))^2 \right)} e^{-\jj 2\pi \fD \tau \cos(\alpha) } \rmd \alpha .
\end{align}
The exponent can be simplified as follows.
\begin{align}
(\alpha-\mu_\rmr)^2+(\alpha-(\mu_\rmr + \Delta_\mu))^2  & = \left( \alpha - \left( \mu' - \frac{\Delta_\mu}{2} \right) \right)^2 + \left( \alpha - \left( \mu' + \frac{\Delta_\mu}{2} \right) \right)^2 \nonumber \\
\label{eq:exponent_calc}
& = 2(\alpha-\mu')^2 + \frac{\Delta_\mu^2}{2}.
\end{align}
Substituting this into \eqref{eq:NLOS_corr_gaussian_approx} and approximate $\mu'\simeq \mu$, which is valid for small $\Delta_\mu$, we have
\begin{align}
R_{\rmnlos} (\tau) \simeq \int_{-\pi}^{\pi} \frac{1}{\sqrt{2\pi/\kr}}e^{\frac{\kr}{2}(\alpha-\mu_\rmr)^2}e^{-\frac{\kr\Delta_\mu^2}{8}} e^{-\jj 2\pi \fD \tau \cos(\alpha) } \rmd \alpha. 
\end{align}
To obtain a final closed form expression, the Gaussian PDF is approximated back to von Mises one to get the following.
\begin{align}
R_{\rmnlos} (\tau) & \simeq \frac{e^{ - \frac{\kr \Delta_\mu^2}{8} }}{2\pi I_0(\kr)} \int_{-\pi}^{\pi} e^{\kr\cos(\alpha-\mu_\rmr) } e^{-\jj 2\pi \fD \tau \cos(\alpha) } \rmd \alpha \nonumber \\
 \label{eq:NLOS_corr_func_appr}
 & = e^{ - \frac{\kr \fDsq\tau^2\sin^2\mu_\rmr}{8 D_{\rmr,\lambda}^2} } \frac{I_0(\sqrt{x^2+y^2})}{I_0(\kr)}
\end{align}
where, 
\begin{align}
\label{eq:x}
x & = k_\rmr\cos\mu_\rmr-\jj2\pi \fD \tau,  \\
\label{eq:y}
y & = k_\rmr\sin\mu_\rmr.
\end{align}
In this \redtext{paper}, we are interested in narrow receive beamwidths (i.e., $\kr$ large), and this approximation turns out to be decent enough for our purpose as will be shown in the numerical examples at the end of this section. 
Note that in the approximation in \eqref{eq:NLOS_corr_func_appr}, the effect of pointing error due to the receiver motion is decoupled from the usual effect of Doppler spread to the channel. 

\subsection{LOS Channel Correlation Function} \label{sec:corr_func_los}
The correlation function for the LOS channel is defined as
\begin{align}
R_{\rmlos}(\tau)=\frac{1}{\max\curlbr{|h_\rmlos(t)|^2,|h_\rmlos(t+\tau)|^2}}h_{\rmlos}(t) h_{\rmlos}^*(t+\tau),
\end{align}
where the normalization is to ensure that $|R_{\rmlos}(\tau)|\le 1$. 
Substituting the channel in \eqref{eq:channel_model_LOS} we have
\begin{align} 
\label{eq:corr_los_gen}
R_{\rmlos}(\tau) = & \frac{\sqrt{G(\alpha_{\rmlos}| \mu_\rmr)G(\alpha_{\rmlos}| \mu_\rmr+\Delta_\mu^\rmlos)}}{\max\curlbr{G(\alpha_{\rmlos}| \mu_\rmr),G(\alpha_{\rmlos}| \mu_\rmr+\Delta_\mu^\rmlos)}}  \nonumber \\ 
& \times e^{\jj 2\pi \fD t [\cos(\alpha_{\rmlos}) - \cos(\alpha_{\rmlos}+\Delta_\mu^\rmlos)]} e^{-\jj 2\pi \fD \tau \cos(\alpha_{\rmlos}+\Delta_\mu^\rmlos)}, 
\end{align}
where we have incorporated the receive beam pointing error due to the receiver motion over the time period $\tau$ through $\Delta_\mu^\rmlos$ as given in \eqref{eq:point_ang_diff_los}. 
Note that \eqref{eq:corr_los_gen} depends on $t$ and thus is not wide sense stationary. In the case of small $\Delta_\mu^\rmlos$, it can be approximated as wide sense stationary as the term $e^{\jj 2\pi \fD t [\cos(\alpha_{\rmlos}) - \cos(\alpha_{\rmlos}+\Delta_\mu^\rmlos)]}\simeq 1$.  
Note that $|R_{\rmlos}(\tau)|= 1$ only when $\Delta_\mu^\rmlos=0$.

If we assume that at time $t$ the receive beam is pointing at $\alpha_\rmlos$, then $G(\alpha_{\rmlos}| \mu_\rmr=\alpha_{\rmlos})=e^\kr/(2\pi I_0(\kr))$ and $G(\alpha_{\rmlos}| \mu_\rmr+\Delta_\mu^\rmlos)=e^{\kr\cos(\Delta_\mu^\rmlos)}/(2\pi I_0(\kr))$ and we have
\begin{align}
\label{eq:corr_los1}
R_{\rmlos}(\tau) 
&=\sqrt{e^{\kr\pa*{\cos(\Delta_\mu^\rmlos)-1}}} e^{\jj 2\pi \fD t [\cos(\alpha_{\rmlos}) - \cos(\alpha_{\rmlos}+\Delta_\mu^\rmlos)]} e^{-\jj 2\pi \fD \tau \cos(\alpha_{\rmlos}+\Delta_\mu^\rmlos)} \\
\label{eq:corr_los2}
 & \simeq e^{\frac{1}{2}\kr\pa*{\cos(\Delta_\mu^\rmlos)-1} } e^{-\jj 2\pi \fD \tau \cos(\alpha_{\rmlos})},
\end{align}
where the approximation follows when $\Delta_\mu^\rmlos$ is small, which typically is the case because the transmitter-receiver distance $D$ is large. Taking the absolute value of either \eqref{eq:corr_los1} or \eqref{eq:corr_los2} gives
\begin{align}
\label{eq:corr_los3}
|R_{\rmlos}(\tau)| & = e^{\frac{1}{2}\kr\pa*{\cos(\Delta_\mu^\rmlos)-1} }.
\end{align}
The expression in \eqref{eq:corr_los3} \redtext{means} that the only factor affecting the channel correlation of the LOS channel is the pointing error. 

\subsection{Numerical Verification of \eqref{eq:NLOS_corr_func_appr} and Effect of K Factor}
\begin{figure}
\centering
\includegraphics[width=0.6\textwidth]{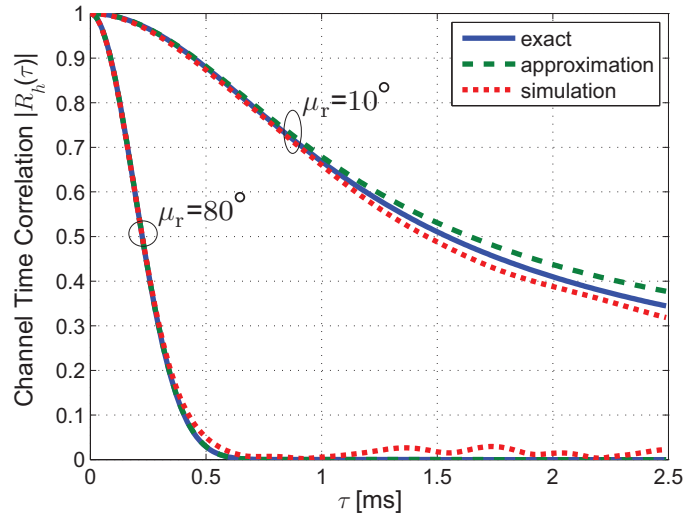}
\caption{Comparison of the channel \redtext{temporal} correlations computed from the exact expression \eqref{eq:NLOS_corr_func_exact}, the approximate expression \eqref{eq:NLOS_corr_func_appr}, and those computed from simulation. The beamwidth is fixed at around $8^\circ$ ($\kr=50$), 
receiver speed $v=30$ m/s, the carrier frequency $\fc=60$ GHz, 
and the scattering radius $D_{\rmr,\lambda}=100$ were used.
}
\label{fig:corr_kr=50_Dr=100}
\end{figure}

First we will verify our approximation for the NLOS case in \eqref{eq:NLOS_corr_func_appr} by comparing it with the exact expression given in \eqref{eq:NLOS_corr_func_exact} and the correlation computed from simulation. \redtext{Then we provide an example to show the effect of the K factor on the channel correlation.} The derivation of \eqref{eq:NLOS_corr_func_appr} is based on the assumption of small $\Delta_\mu$, which depends on both the scattering radius $D_{\rmr,\lambda}$ and the pointing direction $\mu_\rmr$. This assumption is most restrictive when $\mu_\rmr$ approaches $\pi/2$ and $D_{\rmr,\lambda}$ small. With this in mind, we set 
the transmitter-receiver distance $D=50$ m, the receiver speed $v=30$ m/s, the carrier frequency $f_\mathrm{c}=60$ GHz, the scattering radius $D_{\rmr}=0.5$ m (i.e., $D_{\rmr,\lambda}=100$). We compute the the case when $\mu_\rmr=10^\circ$ and when $\mu_\rmr=80^\circ$ to compare the effect of $\mu_\rmr$. Note that there was no assumption on the receive beamwidth in the derivation, and the accuracy of this approximation does not depend on the beamwidth. For this reason we fix $\kr=50$, which corresponds to a beamwidth of around $8^\circ$. We simulate the channel following the sum of sinusoid approach \cite{Zajic:Space-time-correlated-mobile-to-mobile:08} using the model given in \eqref{eq:channel_model_NLOS}. Note that the transmitter-receiver distance $D$ is used only in the simulation and is not used in the exact or the approximate expression given in \eqref{eq:NLOS_corr_func_exact} and \eqref{eq:NLOS_corr_func_appr}. As can be seen in Fig. \ref{fig:corr_kr=50_Dr=100}, all the three curves match very well. The deviation of the simulation curve from the exact solution is due to the slow convergence of the sum of sinusoid approach (10000 sinusoids were summed in the simulation).   

\begin{figure} 
\centering
\subfigure[When $\mu_\rmr=10^\circ$]{%
\includegraphics[width=0.47\columnwidth]{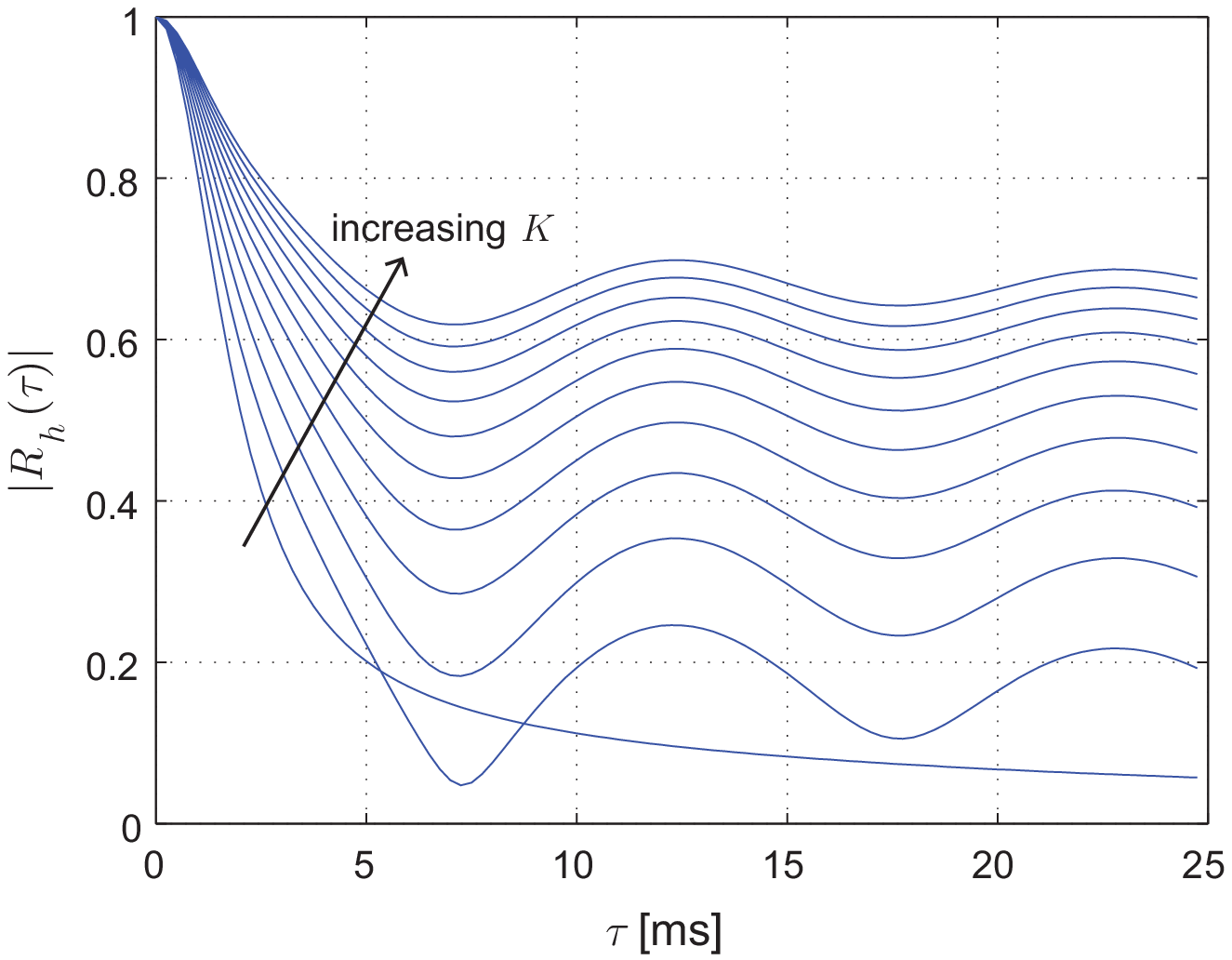}
\label{fig:sig_v=0_01}
}
\quad
\subfigure[When $\mu_\rmr=80^\circ$]{%
\includegraphics[width=0.47\columnwidth]{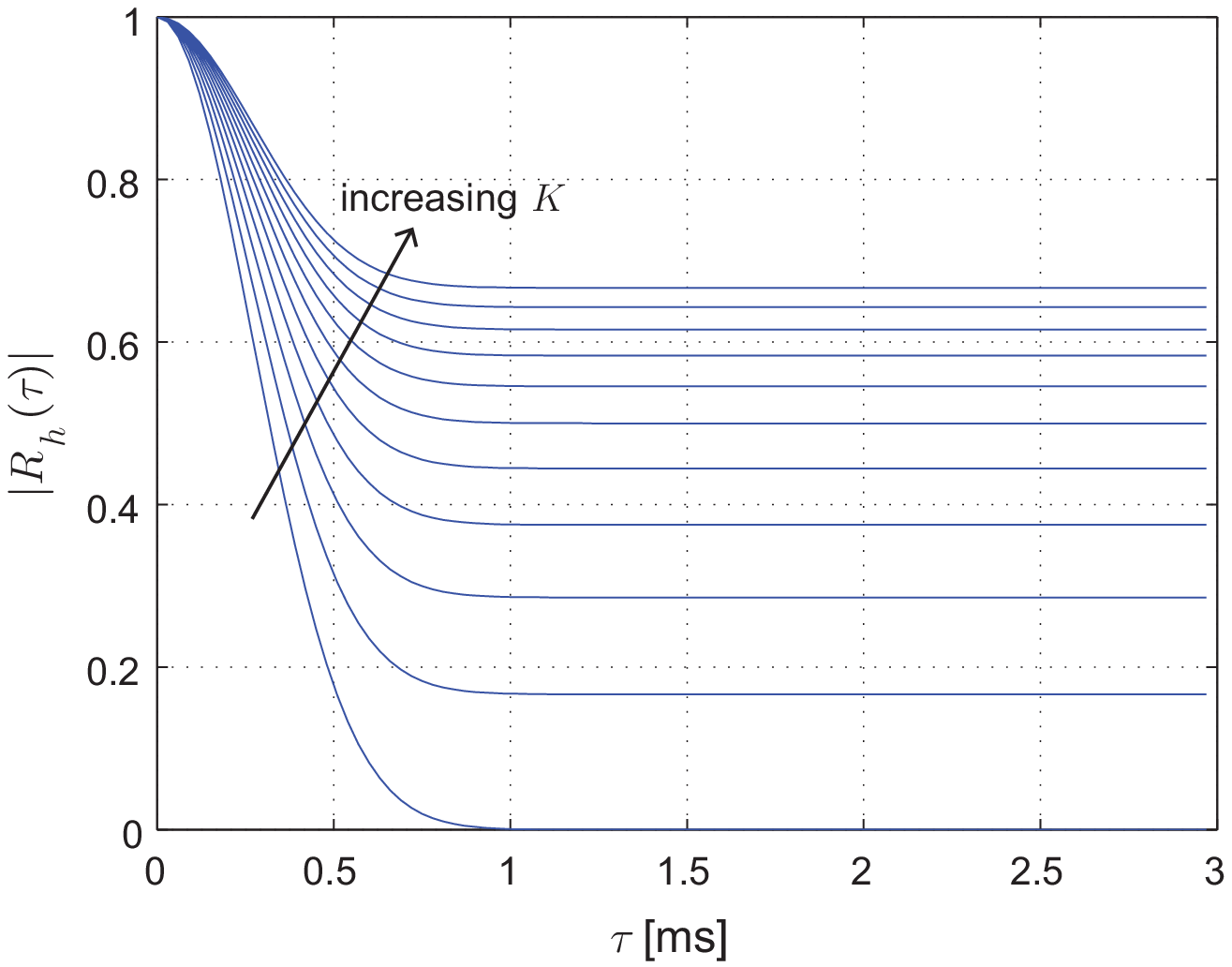}
\label{fig:sig_v=0_1}
}
\caption{The effect of the Rician K factor on the channel correlation function. The plots show the absolute values of the correlation coefficients for $K$ from $0$ to $2$ with a step of $0.2$.}
\label{fig:Rh_kr=50_Dr=1000_D=50}
\end{figure}

Since the derivation of the correlation for the LOS channel is simple, we will skip its verification. Instead we provide an example showing how the $K$ factor affects the channel correlation as shown in Fig. \ref{fig:Rh_kr=50_Dr=1000_D=50}. The parameters are the same as in the NLOS case  except $D_{r,\lambda}=1000$. 
The channel correlation oscillates for $\mu_\rmr=10^\circ$ case but it is not observed for the $\mu_\rmr=80^\circ$ case. The oscillation is due to the phase difference between the LOS component $R_\rmlos(\tau)$ and the NLOS component $R_\rmnlos(\tau)$. When $\mu_\rmr$ is close to $90^\circ$, the $R_\rmnlos(\tau)$ decreases fast (note that $\mu_\rmr=90^\circ$ corresponds to the fastest fading case) so that before the phase difference between $R_\rmlos(\tau)$ and $R_\rmnlos(\tau)$ takes effect, the NLOS component $R_\rmnlos(\tau)$ decreases to a negligible value in relation to $R_\rmlos(\tau)$ and the oscillation is not observed. 
It can also be observed from the plots that the channel correlation increases with $K$ regardless of $\mu_\rmr$. 
This is because $\Delta_{\mu}^\rmlos$ increases very slowly with $\tau$ (because typically $D_\lambda$ is very larger, and in particular in this example $D_\lambda=10^4$) which leads to very slow decrease in the LOS correlation component $|R_\rmlos(\tau)|$ compared to that of the NLOS component $|R_\rmnlos(\tau)|$. 


\section{Channel Coherence Time} \label{sec:chan_coherence_time}
Using the result from 
the previous section, the coherence times are derived in this section. The channel coherence time is defined as the time $\tau=\Tc$ at which the channel correlation decreases to $|R_h(\tau)|=R$ for some predefined value $R$. Typically $R$ ranges from $0.3$ to $0.9$ \cite{Medard2000}. Note that for a given channel, requiring a larger $R$ will result in a smaller $\Tc$. 
A general solution to $|R_h(\tau)|=R$ is intractable because both $R_\rmlos(\tau)$ and $R_\rmnlos(\tau)$ are complex numbers and $R_\rmnlos(\tau)$ is a complicated function involving the Bessel function. Instead of dealing with this directly, 
we will derive the coherence time for the NLOS and LOS case separately, which provide insights for the upper and lower bound on the channel coherence time.

\subsection{NLOS Channel}
The coherence time expressions for the case when only the NLOS component exists are derived using the correlation function given in \eqref{eq:NLOS_corr_func_appr}. Due to the Bessel function, the solution for a general main beam direction $\mu_\rmr$ and arbitrary beamwidth $\theta$ is intractable. In the following we assume small $\theta$ and we will solve for two cases, namely when $|\mu_\rmr|$ is small and when $|\mu_\rmr|$ is not small. Note that the case where $\mu_\rmr=0$ is when the main beam direction is parallel to the direction of travel resulting the slowest fading case, while $|\mu_\rmr|=\pi/2$ is when the main beam direction is perpendicular to the direction of travel and the receiver will experience the fastest fading \cite{Durgin2000}. For most cases, our approximation is valid for beamwidth $\theta$ up to around $20^\circ$. 
\greentext{This is not a big limitation because most likely mmWave systems will use narrow beams. For example, a prototype system developped by Samsung Electronics uses an array with $10^\circ$ beamwidth in azimuth \cite{Roh2014} and long range automotive radars use beamwidth on the order of a few degrees \cite{Wenger2005}.}


\subsubsection{When $|\mu_\rmr|$ is small} For small $\mu_\rmr$, we approximate $x\simeq \kr - \jj 2\pi \fD\tau$ and $y\simeq 0$, where $x$ and $y$ are defined in \eqref{eq:x} and \eqref{eq:y}, respectively. Note that the accuracy of this approximation depends on both $\kr$ and $\mu_\rmr$. 
When $\mu_\rmr$ is small, $y\simeq \kr \mu_\rmr \simeq \mu_\rmr/\theta^2$ and thus roughly the approximation works for $\theta>\sqrt{\mu_\rmr}$. Assuming $\theta$ is in this range we can apply the approximation to get
\begin{align}
R_h(\tau) &\simeq e^{ - \frac{\kr \fDsq\tau^2\sin^2\mu_\rmr}{8 D_{\rmr,\lambda}^2} } \frac{I_0( \kr - \jj 2\pi \fD\tau  )}{I_0(\kr)} \\
\label{eq:corr_small_mu_appr}
& \simeq e^{ - \frac{\kr \fDsq\tau^2\sin^2\mu_\rmr}{8 D_{\rmr,\lambda}^2} } \frac{e^{ -\jj 2\pi \fD \tau}}{\sqrt{1-\jj 2\pi \fD\tau / k_\rmr}}. 
\end{align}
The last step follows by applying the asymptotic approximation of the Bessel function \cite{Abramowitz:Handbook-of-mathematical-function:74}
\begin{align}
\label{eq:Bessel_appr}
I_0(z) \simeq \frac{e^z}{\sqrt{2\pi z}},
\end{align}
which holds for $|z|$ large. Taking the absolute value to get
\begin{align}
|R_h(\tau)| &\simeq e^{ - \frac{\kr \fDsq\tau^2\sin^2\mu_\rmr}{8 D_{\rmr,\lambda}^2} } \frac{1}{|\sqrt{1-\jj 2\pi \fD\tau / k_\rmr}|} \\
\label{eq:Tc_eq_nlos_small_mu}
&= e^{ - \frac{\kr \fDsq\tau^2\sin^2\mu_\rmr}{8 D_{\rmr,\lambda}^2} } \frac{1}{(1 + (2\pi \fD\tau / k_\rmr)^2)^{1/4} }.
\end{align}
Following the definition $|R_h(\Tc)|=R$, we can solve for $\Tc$ as follows.
\begin{align}
1 + (2\pi \fD\Tc / k_\rmr)^2 & = \frac{1}{R^4} e^{ - \frac{\kr \fDsq T_\rmc^2\sin^2\mu_\rmr}{2 D_{\rmr,\lambda}^2} }
\\ & \simeq \frac{1}{R^4} \left( 1  - \frac{\kr \fDsq T_\rmc^2\sin^2\mu_\rmr}{2 D_{\rmr,\lambda}^2}  \right) \\
\Rightarrow \Tc & = \sqrt { \frac{1/R^4 -1}{ (2\pi \fD / k_\rmr)^2 + \frac{\kr \fDsq\sin^2\mu_\rmr}{2 D_{\rmr,\lambda}^2R^4 }  } }
\end{align}
where we have used the approximation $e^z\simeq 1+z$ to eliminate the exponential term. 
For small beamwidth we have $\kr\simeq 1/\theta^2$ and thus in terms of beamwidth we have
\begin{align}
\label{eq:Tc_nlos_small_mu}
\Tc(\theta) & = \sqrt { \frac{1/R^4 -1}{ (2\pi \fD )^2\theta^4 + \frac{1}{2 \theta^2R^4} \left(\frac{\fD\sin\mu_\rmr}{D_{\rmr,\lambda}}\right)^2  } }.
\end{align}
When $D_{\rmr,\lambda}\to \infty$, i.e., ignoring the pointing angular change due to the receiver movement, the coherence time simplifies to
\begin{align}
\label{eq:Tc_nlos_small_mu_no_ang_diff}
\Tc(\theta) & = \frac{\sqrt{1/R^4 -1 }}{2\pi \fD \theta^2}.
\end{align}
We can see that in this case the coherence time is inversely proportional to $\theta^2$. 

\subsubsection{When $|\mu_\rmr|$ is not small} The approach here is different from the previous case. First we compute the argument of the Bessel function, and then we apply asymptotic approximation \eqref{eq:Bessel_appr}. Taking the $\log$ of the obtained equation, we get a polynomial equation of $\tau$. Exact solution is not trivial, but considering the range of values of the parameters, higher order terms are negligible and we can approximately solve a quadratic equation instead. 

Defining $c+\jj d = \sqrt{x^2 + y^2}$ where $x$ and $y$ are given in \eqref{eq:x} and \eqref{eq:y} respectively. With simple algebra, $c$ and $d$ can be derived as
\begin{align}
c = \sqrt{\frac{\sqrt{a^2+b^2}+a}{2}}, \hspace{1cm} d = \frac{b}{2c},
\end{align}
where $a=\krsq-(2\pi\fD)^2\tau^2$ and $b=-4\pi\fD \kr \cos(\mu_\rmr) \tau$. Substitute $c$ and $d$ above into \eqref{eq:NLOS_corr_func_appr}, apply the asymptotic approximation \eqref{eq:Bessel_appr}, and finally take the absolute value we have
\begin{align}
R = e^{ - \frac{\kr f_{\mathrm{D}}^2 \tau^2\sin^2\mu_\rmr}{8 D_{\rmr,\lambda}^2} } \frac{e^{c-\kr}}{(1+(b/2\krsq)^2)^{1/4}}.
\end{align}
For large $\kr$, the denominator takes values close to one, and 
we approximate $(1+(b/2\krsq)^2)^{1/4}\simeq 1$. Taking the $\log$ on both sides and rearranging we have
\begin{align*}
\kr + \log R + \frac{\kr f_{\mathrm{D}}^2 \tau^2\sin^2\mu_\rmr}{8 D_{\rmr,\lambda}^2} = \sqrt{\frac{\sqrt{a^2+b^2}+a}{2}}.
\end{align*} 
Now taking the square of both sides and ignore the $\tau^4$ term to get
\begin{align*}
2(\kr+\log R)^2 + 4(\kr+\log(R)) & \frac{\kr f_{\mathrm{D}}^2 \tau^2\sin^2\mu_\rmr}{8 D_{\rmr,\lambda}^2} - a= \sqrt{a^2+b^2}.
\end{align*}
Once again taking the square of both sides to get rid of the square root, and neglect the higher order terms with respect to $\tau$. Then substitute $a$ and $b$ we obtain \eqref{eq:Tc_not_small_final_equation}, from which the approximate channel coherence time expression \eqref{eq:Tc_not_small_approx_gen} can be readily derived.   
\begin{align}
 \label{eq:Tc_not_small_final_equation}
 4(\kr+\log R)^4 -4\krsq(\kr+\log R)^2 + \sqbr*{16(\kr+\log R)^3 -8\krsq(\kr+\log R) } \frac{\kr f_{\mathrm{D}}^2 \sin^2\mu_\rmr}{8 D_{\rmr,\lambda}^2}\tau^2 \nonumber \\
 + 4(\kr+\log R)^2(2\pi\fD)^2 \tau^2 = (4\pi\fD\kr\cos\mu_\rmr)^2\tau^2 \\
 \label{eq:Tc_not_small_approx_gen}
 T_{\mathrm{c}}^2 = \frac{\krsq-(\kr+\log R)^2}{ \sqbr*{4(\kr+\log R) -2\frac{\krsq}{\kr+\log R} } \frac{\kr f_{\mathrm{D}}^2 \sin^2\mu_\rmr}{8 D_{\rmr,\lambda}^2} + (2\pi\fD)^2 - \frac{(4\pi\fD\kr\cos\mu_\rmr)^2}{(\kr+\log R)^2} }.
\end{align}
Note that for a fixed $\mu_\rmr$ the denominator in \eqref{eq:Tc_not_small_approx_gen} can be negative leading to invalid solution. The range of valid solution increases with $\mu_\rmr$ as will be shown in our numerical example. As evident from Fig. \ref{fig:Tc_nlos_not_small_mu_r}, if $\mu_\rmr$ is not too small, our result covers most of the beamwidths of interest for mmWave systems.

For the special case when $\mu_\rmr=90^\circ$, which is the fastest fading case, \eqref{eq:Tc_not_small_approx_gen} can be simplified by approximating $4(\kr+\log R) -2\frac{\krsq}{\kr+\log R} \simeq 2(\kr+\log R)$, which is valid for large $\kr$. Finally using the relationship $\kr=\frac{1}{\theta^2}$, the worst case channel coherence time can be expressed as
\begin{align}
\label{eq:Tc_nlos_90}
\Tc(\theta) = \sqrt{\frac{1- \left( 1+\theta^2 \log R \right)^2}{ \frac{1}{4}\left( 1+\theta^2\log R \right)\left(\frac{\fD\sin\mu_\rmr}{D_{\rmr,\lambda}}\right)^2 + (2\pi \fD)^2 \theta^4  }}.
\end{align}

When $D_{\rmr,\lambda}\to \infty$, this simplifies to
\begin{align}
\Tc(\theta) = \frac{1}{2\pi \fD}\sqrt{\frac{1}{\theta^2}\log \frac{1}{R^2}- \left( \log R \right)^2 }.
\end{align}
Using the approximation $\sqrt{1+z}\simeq 1+\frac{1}{z}$ for small $z$, it can be shown that $\Tc(\theta)$ increases on the order of $1/\theta$ for small $\theta$ at the pointing angle $\mu_\rmr=90^\circ$.

\subsection{LOS Channel}
When the LOS dominates, $K/(K+1)\to 1$ and  $R_h(\tau)\simeq R_\rmlos(\tau)$. Thus we have
\begin{align}
|R_h(\tau)| = e^{\frac{1}{2}\kr\pa*{\cos(\Delta_\mu^\rmlos)-1} }.
\end{align} 
Substituting \eqref{eq:point_ang_diff_los} for $\Delta_\mu^\rmlos$ and using the definition $|R_h(\Tc)|=R$, the above equation can be easily solved to get
\begin{align}
\Tc(\theta) & = \frac{D_\lambda}{\fD \sin(\alpha_\rmlos)}\cos^{-1}\pa*{\frac{2}{\kr}\log R+1} \\
\label{eq:Tc_los}
& = \frac{D_\lambda}{\fD \sin(\alpha_\rmlos)}\cos^{-1}\pa*{2\theta^2\log R+1}.
\end{align}
Note that for this expression to be computable $\frac{2}{\kr}\log R+1 \in [-1,1]$ must hold. Within the range of interest for $R$, typically within $[0.3,1]$, $\frac{2}{\kr}\log R+1 \in [-1,1]$ is true for all $\kr>2$. Since we are interested in large $\kr$, this constraint presents no limitation here.  

\subsection{Numerical Results}
We will provide numerical result to verify the derivation for the NLOS case. The derivation for the LOS case is straightforward and thus no verification is given here. 
The following parameters are used: the receiver speed $v=30$ m/s, the carrier frequency $\fc=60$ GHz ($\lambda=5$ mm), the scattering radius $D_{\rmr,\lambda}=100$ and the target correlation $R=0.5$. As the approximations depend on $\mu_\rmr$, we investigate their behavior for different values of $\mu_\rmr$ in the followings.

\begin{figure} 
\centering
\subfigure[$\mu_\rmr=1^\circ$]{%
\includegraphics[width=0.47\columnwidth]{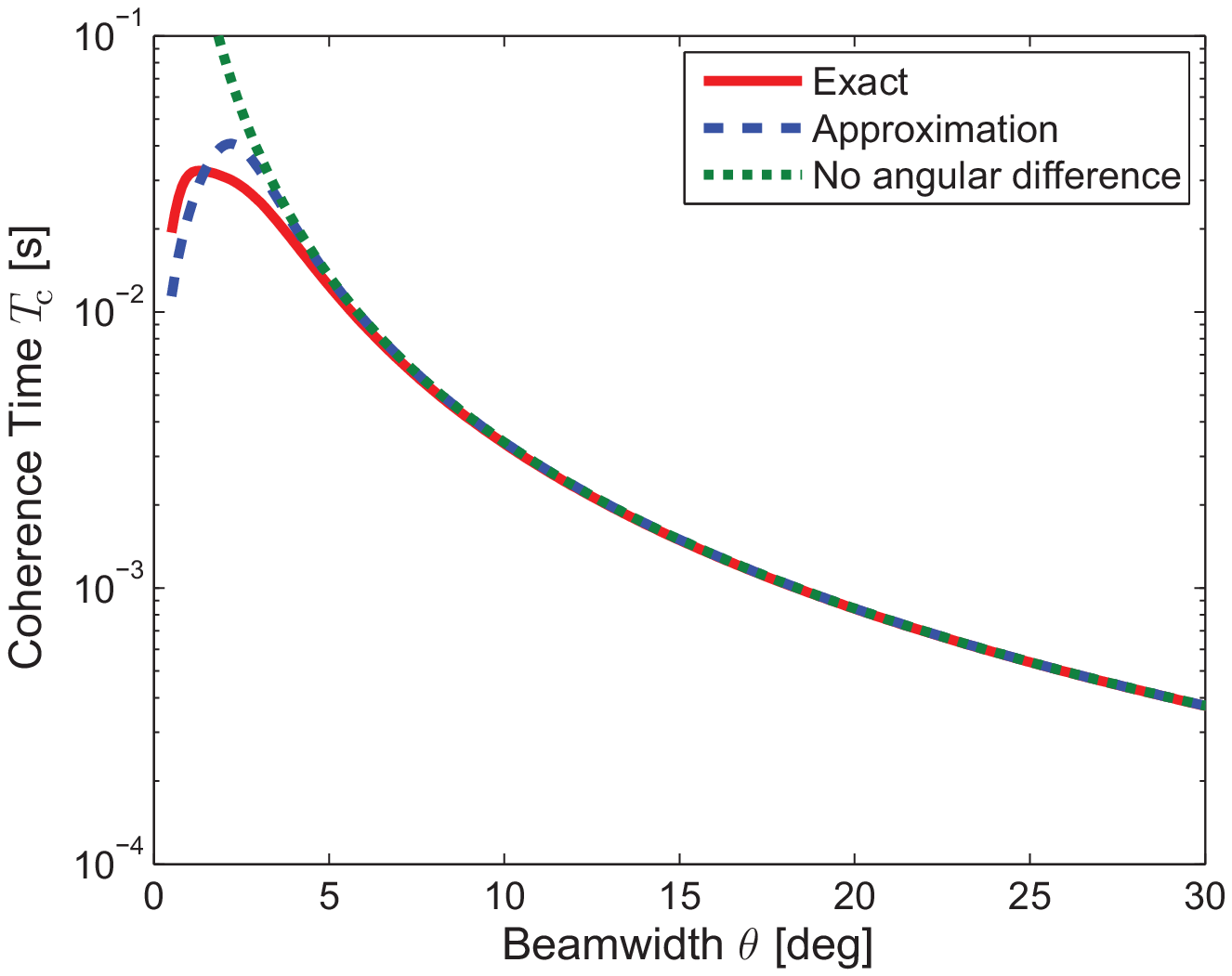}
\label{fig:mu_r=1}
}
\quad
\subfigure[$\mu_\rmr=5^\circ$]{%
\includegraphics[width=0.47\columnwidth]{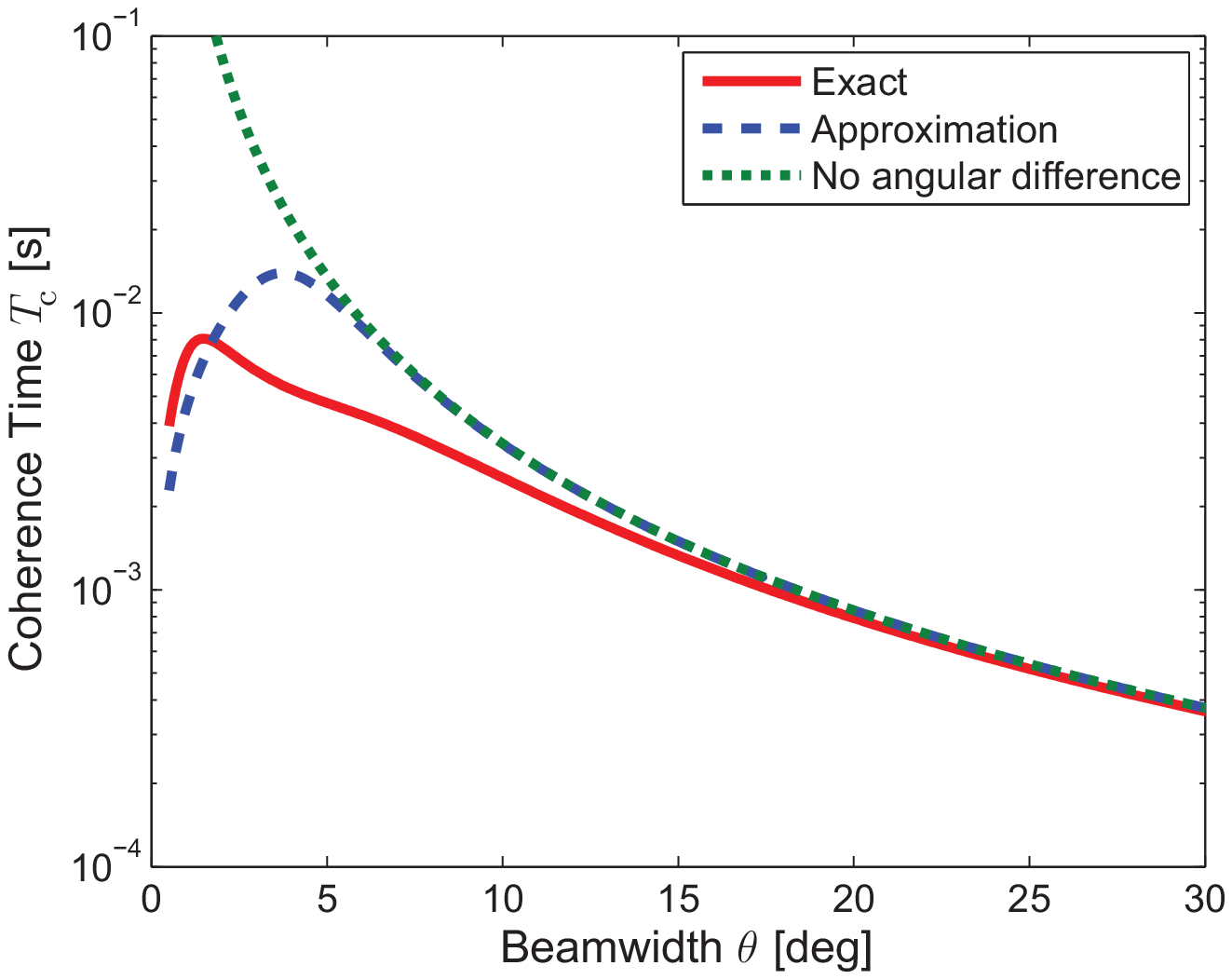}
\label{fig:mu_r=5}
}
\caption{Channel coherence time for small $\mu_\rmr$. ``Exact" refers to the coherence time obtained numerically from the exact correlation function \eqref{eq:NLOS_corr_func_exact}. ``Approximation" and ``No angular difference" refer to \eqref{eq:Tc_nlos_small_mu} and \eqref{eq:Tc_nlos_small_mu_no_ang_diff}, respectively. The result is quite sensitive to $\mu_\rmr$, and the approximation does not work well for small $\theta$ but still can capture the effect of the receiver motion.}
\label{fig:Tc_small_mu_r}
\end{figure}

For the small $|\mu_\rmr|$ case, to see the sensitivity of the approximation in \eqref{eq:Tc_nlos_small_mu} we plot the expression and compare it with that of the exact solution for $\mu_\rmr=1^\circ$ and $\mu_\rmr=5^\circ$.  The exact solution is obtained numerically using the exact correlation function \eqref{eq:NLOS_corr_func_exact}. The ``Approximation" and ``No angular difference" refers to the expressions in \eqref{eq:Tc_nlos_small_mu} and \eqref{eq:Tc_nlos_small_mu_no_ang_diff}, respectively. As mentioned in the derivation, for a given $\mu_\rmr$ the approximation does not work well for $\theta$ too small. This error becomes more severe when $|\mu_\rmr|$ gets larger, which can be seen in the plots in Fig. \ref{fig:Tc_small_mu_r}. Nevertheless, the result can still capture the effect of the receiver motion. 

\begin{figure} 
\centering
\includegraphics[width=0.6\columnwidth]{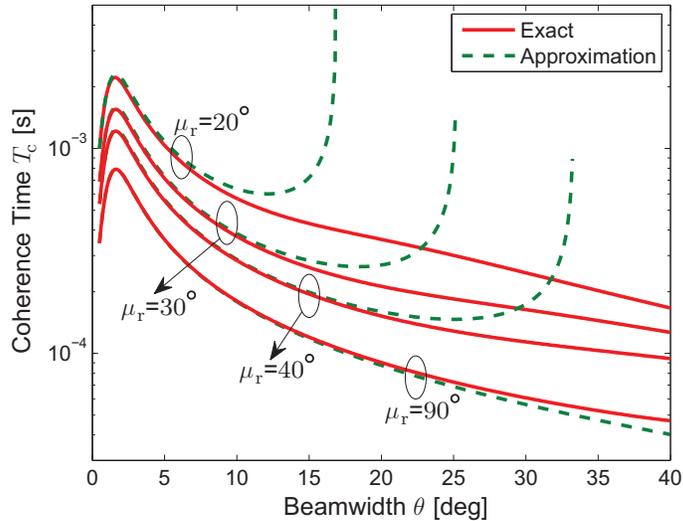}
\caption{Channel coherence time for difference values of $\mu_\rmr$. ``Exact" refers to the numerical solution to the exact correlation function \eqref{eq:NLOS_corr_func_exact}, and ``Approximation" refers to \eqref{eq:Tc_not_small_approx_gen}. It can be seen that the range of valid approximation increases with $\mu_\rmr$.}
\label{fig:Tc_nlos_not_small_mu_r}
\end{figure}

The same study is done for the case when $\mu_\rmr$ is not small. The result is shown in Fig. \ref{fig:Tc_nlos_not_small_mu_r} for four different values of $\mu_\rmr$. For a fixed value of $\mu_\rmr$ there is a point where the approximation diverges due to the singularity of the denominator in \eqref{eq:Tc_not_small_approx_gen}. We can also observe that the range of $\theta$ for valid approximation increases with $\mu_\rmr$, and it is valid up to around $\mu_\rmr/2$, i.e., half the pointing angle. Considering the fact that mmWave systems require narrow beams to compensate for the high path loss, this channel coherence time approximation will be valid for most cases of interest in practice.

\section{Beam Coherence Time} \label{sec:beam_coherence_time}
This section first defines the beam coherence time, and then the beam coherence time expressions are given for the LOS and NLOS cases. The NLOS case will be based on our spatial lobe model described in Section \ref{sec:chan_spatial_lobe_model}. 

The beam coherence time is defined as the average time over which the beam stays aligned. We focus on only the receive beam here. 
For a given receive beamwidth, the beam is said to become misaligned when the receive power falls below a certain ratio $\zeta\in[0,1]$ compared to the peak receive power. 
Assuming that the receive beam is pointing at the peak direction $\mu_\rmr$ at time $t$, then we can define the beam coherence time by
\begin{align}
\label{eq:misaligned_cond_time}
\TB = \inf_{\tau} \left\{ \tau \left|  \frac{P(t+\tau)}{P(t)} < \zeta   \right. \right\}.
\end{align}
Note that the power decrease here is due to the pointing error $\Delta_\mu(\tau)$ as defined in Section \ref{sec:pointing_error_model}. Using this definition, we derive the beam coherence times 
in the followings.

\subsection{LOS Case}
The pointing angular change $\Delta_\mu^\rmlos$ is given in \eqref{eq:point_ang_diff_los}. Let the beam pattern be represented by the von Mises distribution as earlier, the receive power is proportional to the receive beam pattern. In particular, $P(t) \propto G(\mu_\rmr|\mu_\rmr)$ and $P(t+\tau) \propto G(\mu_\rmr|\mu_\rmr+\Delta_\mu^\rmlos)$. At $\tau=\TB$,
\begin{align}
\zeta & = \frac{G(\mu_\rmr|\mu_\rmr+\Delta_\mu^\rmlos)}{G(\mu_\rmr|\mu_\rmr)} \\
& = e^{\kr(\cos(\Delta_\mu^\rmlos) -1)}.
\end{align}
Substituting $\Delta_\mu^\rmlos$ from \eqref{eq:point_ang_diff_los} the above equation can be easily solved to get
\begin{align}
\label{eq:TB_los}
\TB(\theta) = \frac{D_\lambda}{\fD \sin \mu_\rmr} \cos^{-1}\pa*{\theta^2\log \zeta +1},
\end{align}
where we have used $\kr=1/\theta^2$ in the above expression. Note that this expression is closely related to the expression for the channel coherence time for the LOS case. 

\subsection{NLOS Case}
First we need to find the pointing angular change due to the receiver motion. In the NLOS case, the incoming power is the result of the reflection from the scatterers. Following our one-ring scatter model, the pointing angular change is given by \eqref{eq:point_ang_diff_tau}. 

Now we need to compute the receive power and use the expression to solve for $\TB$. Assuming the spatial lobes are also represented by the von Mises distribution with lobe width given by $\beta$, which is defined in \eqref{eq:lobe_width_dist}, or by the concentration parameter $\kappa_\rmr=1/\beta^2$, and has peak at $\mu_\rmr$. This is the PAS $\mcP'(\alpha|\mu_\rmr)$ before applying the receive beam pattern. 
At time $t$ assume that the receive beam is pointing at the peak of $\mcP'(\alpha|\mu_\rmr)$, i.e., using the beam pattern $G(\alpha|\mu_\rmr)$. At time $t+\tau$, the beam pattern now changes to $G(\alpha|\mu_\rmr+\Delta_\mu)$ if no realignment is done. The receive power at a pointing angle $\mu_\rmr+\Delta_\mu$ is given by
\begin{align}
P(t+\tau) = \int_{0}^{2\pi} \mcP'(\alpha|\mu_\rmr)G(\alpha|\mu_\rmr+\Delta_\mu) \rmd \alpha.
\end{align}
Note that for large $\kr$ the von Mises PDF in \eqref{eq:von_Mises_pdf} approaches the Gaussian PDF \cite[Ch. 45]{Forbes2010}. 
Also note that for large $\kr$, i.e., small variance, the distribution falls off fast and tails at both sides beyond $0$ and $2\pi$ have little weight. These observations lead to the following approximation:
\begin{align}
P(t+\tau) & \simeq \int_{-\infty}^{\infty} \frac{1}{\sqrt{2\pi \beta^2}}e^{-\frac{(\alpha-\mu_\rmr)^2}{2\beta^2}} \frac{1}{\sqrt{2\pi \theta^2}}e^{-\frac{(\alpha-\mu_\rmr-\Delta_\mu)^2}{2\theta^2}} \rmd \alpha \nonumber \\
\label{eq:appr_Gaussian_convol}
& = \int_{-\infty}^{\infty} \frac{1}{\sqrt{2\pi \beta^2}}e^{-\frac{(u)^2}{2\beta^2}} \frac{1}{\sqrt{2\pi \theta^2}}e^{-\frac{(\Delta_\mu-u)^2}{2\theta^2}} \rmd u.
\end{align}
Applying a change of variable $u=\alpha-\mu_\rmr$, $\mu_\rmr$ can be eliminated from the first expression.   
The expression \eqref{eq:appr_Gaussian_convol} is just a convolution between two Gaussian PDFs, which is well-known to result in another Gaussian PDF with mean $\Delta_\mu$ and variance $\beta^2+\theta^2$ \cite{Weisstein}. That is,
\begin{align}
\label{eq:conv_gaussian}
P(t+\tau) \simeq \frac{1}{\sqrt{2\pi (\beta^2+\theta^2)}} e^{-\frac{\Delta_\mu^2}{2(\beta^2+\theta^2)}}, 
\end{align}
which does not depend on $\mu_\rmr$. This makes sense because in the current setting it is assumed that at time $t$ the receive beam is aligned to $\mu_\rmr$ and $P(t+\tau)$ is determined solely from the misalignment that happens at time $t+\tau$. This misalignment is captured by pointing error due to the receiver motion $\Delta_\mu$, which is a function of $\tau$. We can solve for $\TB$ directly from \eqref{eq:conv_gaussian}; however, by approximating \eqref{eq:conv_gaussian} by a von Mises distribution function, the resulting $\TB$ is of the same form for both the LOS and NLOS cases.    
Doing this approximation, \eqref{eq:conv_gaussian} becomes
\begin{align}
P(t+\tau) \simeq \frac{1}{2\pi I_0(1/(\beta^2+\theta^2))} e^{\frac{\cos(\Delta_\mu)}{\beta^2+\theta^2}}.
\end{align}
With the same steps used in the derivation of the beam coherence time for the LOS case, we get the expression for the NLOS case as
\begin{align}
\TB(\theta|\beta) = \frac{D_{\rmr,\lambda}}{\fD \sin \mu_\rmr} \cos^{-1}\pa*{(\beta^2+\theta^2)\log \zeta +1}.
\end{align}
Note that $\beta$ is a random variable and is modeled by the Gaussian distribution in \eqref{eq:lobe_width_dist}. Thus to get the beam coherence time we need to average over $\beta$,
\begin{align}
\TB(\theta) = \bbE_\beta[\TB(\theta|\beta)].
\end{align}
where $\bbE_\beta[\cdot]$ denotes the statistical expectation over $\beta$. The difference to the LOS case is that now $\TB$ depends on the channel through the spatial lobe angular spread parameter $\beta$. 

\subsection{Numerical Results}
\begin{figure} 
\centering
\subfigure[Beam coherence time for LOS case]{%
\includegraphics[width=0.47\columnwidth]{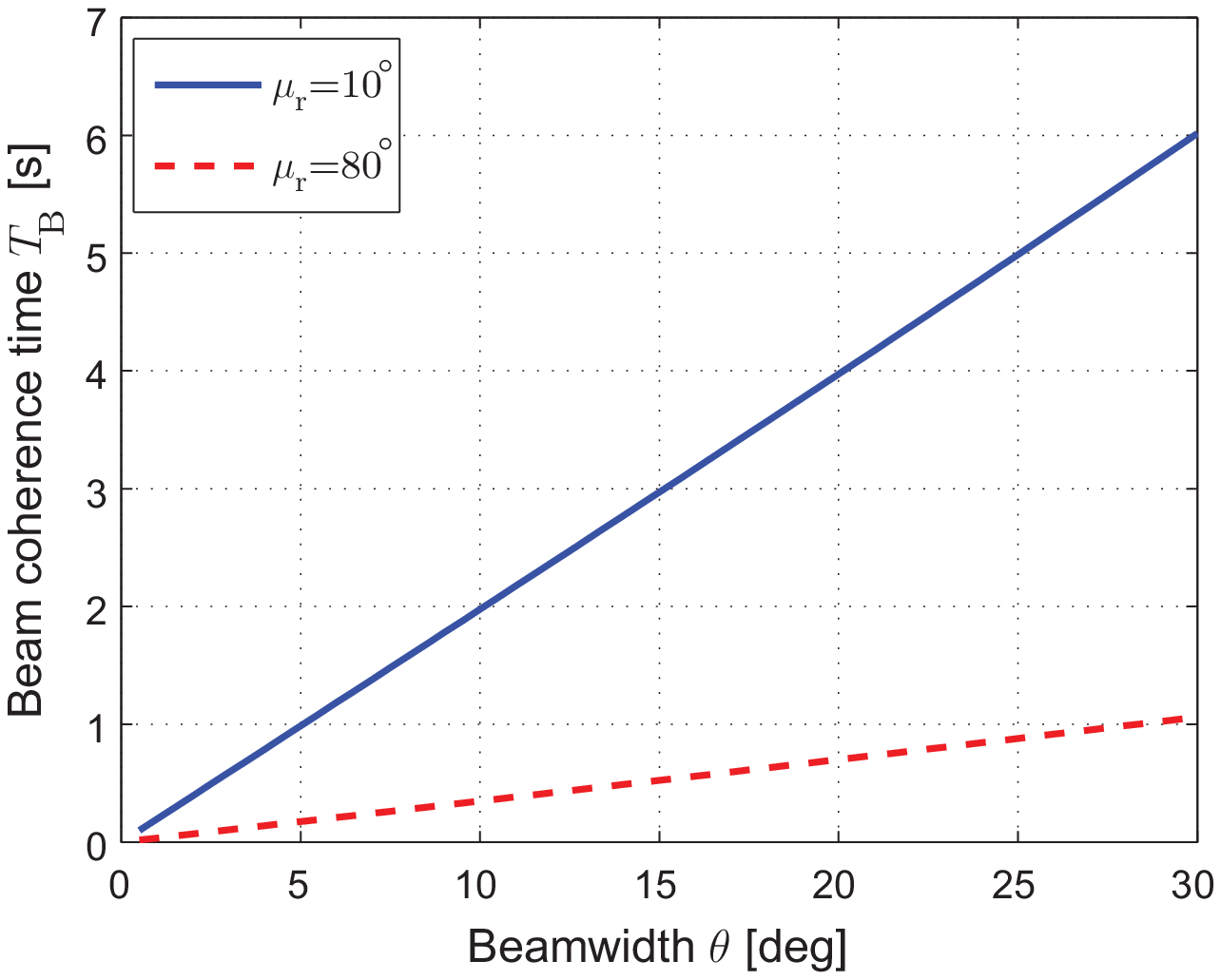}
\label{fig:beam_coherence_LOS}
}
\quad
\subfigure[Beam coherence time for NLOS case]{%
\includegraphics[width=0.47\columnwidth]{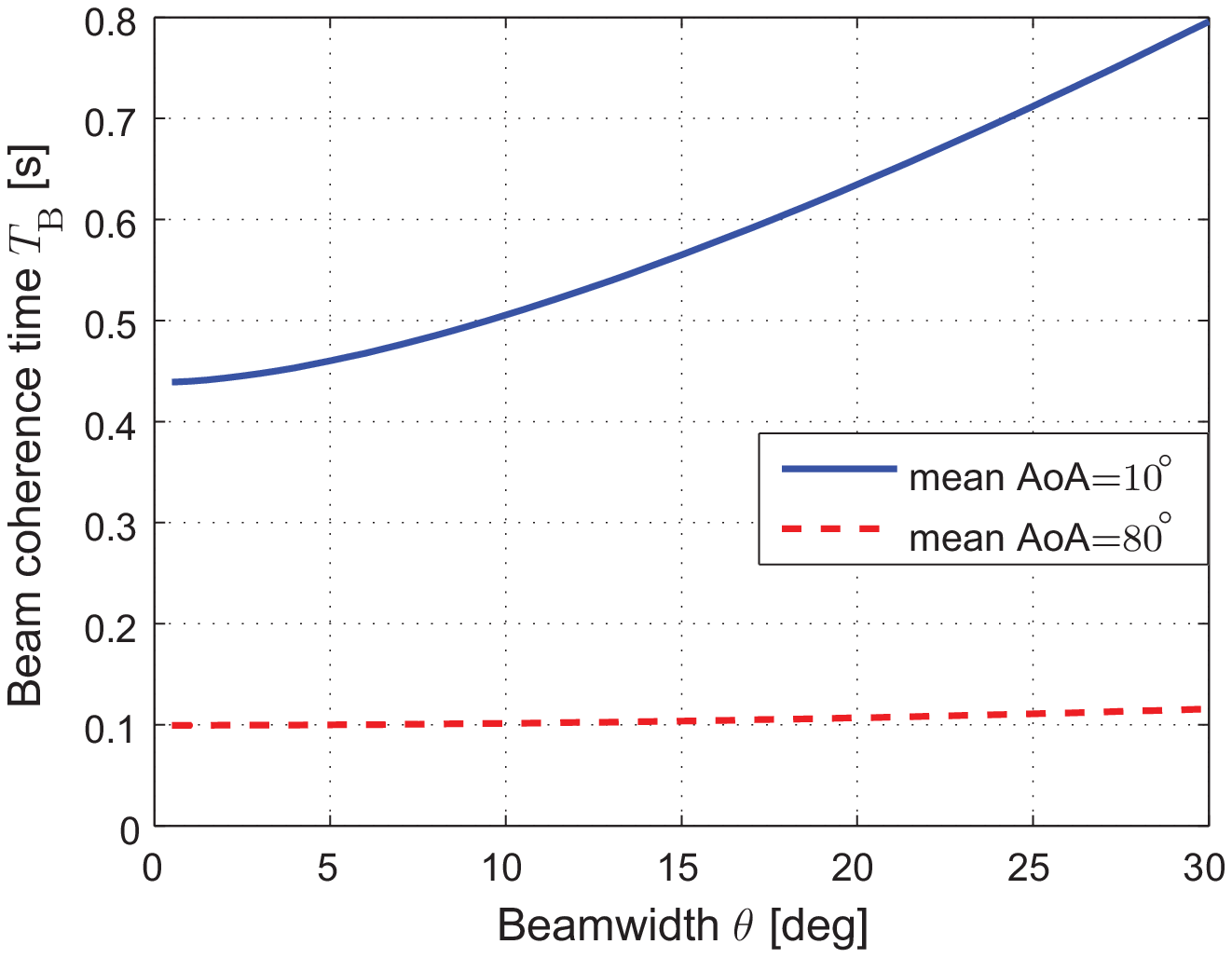}
\label{fig:beam_coherence_NLOS_Drlam=1000}
}
\caption{The beam coherence time for both the LOS and NLOS cases. $v=30$ m/s, $\fc=60$ GHz, and $\zeta=0.5$ are used. For the LOS case, $D=50$ m is used. For the NLOS case $D_{\rmr,\lambda}=1000$ and the angular spread of the angle of arrival is assumed to be Gaussian with standard deviation of $\sigma_{\mathrm{AS}}=25.7^\circ$}
\label{fig:beam_coherence}
\end{figure}

Similar to the previous examples, we set the transmitter-receiver distance $D=50$ m, the receiver speed $v=30$ m/s, and the carrier frequency $\fc=60$ GHz. The threshold power ratio $\zeta$ is set to 0.5. Fig. \ref{fig:beam_coherence_LOS} shows the resulting beam coherence time for the LOS for $\mu_\rmr=10^\circ$ and $\mu_\rmr=80^\circ$. From the plots, it looks like $\TB$ is linear with respect to $\theta$. It is almost linear because the argument to the $\cos^{-1}(\cdot)$ is of the form $1+z^2$, which happens to be the first order Taylor approximation of $\cos(\cdot)$. For the same traveled distance, the pointing angular changes less for small $\mu_\rmr$ which results in larger $\TB$ for $\mu_\rmr=10^\circ$ as shown in the figure.

For the NLOS case, we need statistics of the angular spread of the angles of arrival. Using the result from \cite{Samimi2014}, the angular spread is assumed to be Gaussian with standard deviation $\sigma_{\mathrm{AS}}=25.7^\circ$. The scattering radius $D_{\rmr,\lambda}=1000$ is used and other parameters are as in the LOS case. For the mean angle of arrival of $80^\circ$, we can see that increasing the beamwidth does not effectively increase $\TB$ as in the case when the mean angle of arrival is $10^\circ$.    



\section{Implications on Beam Realignment Duration} \label{sec:implications}
So far we have defined and derived two coherence times: the channel coherence time and the beam coherence time, relevant to scenarios where mmWave directional beams are used in vehicular environments. The channel coherence time determines how fast the channel coefficient changes in time, and thus can be used in deciding the packet length and determining the overhead for channel estimation. We explore some implications of on the choice of beamwidth in this section. 
We argue in Section \ref{sec:realignment_duration} that\redtext{, to maximize the performance,} beam realignment can be done in every beam coherence time not in every channel coherence time. 

\subsection{Lower Bound on Mutual Information}
Consider the following signal model.
\begin{align}
y[i] = h[i]s[i]+n[i], \hspace{1cm} i=1,2,\dots,k
\end{align} 
where $y[i]$ is the receive signal, $h[i]$ is the channel, $s[i]$ is the transmitted signal, $n[i]$ is the additive white Gaussian noise (AWGN), and $k$ is the packet length measured in the number of samples. It is assumed that the channel $h[i]$ is normalized such that its variance is $\sigma_h^2=1$ for all $i$. A simple autoregressive  \redtext{temporal} correlation function is used to describe the time-variation of the channel in discrete-time as 
\begin{align}
h[i] = \alpha h[i-1] + \xi[i-1], \label{eq:gaussMarkov}
\end{align}
where $\alpha$ is the correlation coefficient and is given by 
\begin{align}
\alpha = R_h(\nu T)
\end{align}
with the symbol duration $T$, $\xi[i]$ is the innovation term with variance $\sigma_\xi^2=(1-|R_h(\nu T)|^2)$, and $R_h(\cdot)$ is the channel \redtext{temporal} correlation function derived in Section \ref{sec:corr_func}. 
For decoding, the channel has to be estimated, and the estimation is affected from both the thermal noise and the channel time-variation. 

If the estimator does not have knowledge of the statistics of $\xi[i]$ (which typically is the case), then a natural assumption is that $\xi[i]$ is Gaussian. Because \eqref{eq:gaussMarkov} is a Gauss-Markov channel model, following the logic used in \cite{Medard2000}, the Kalman filter provides the maximum-likelihood estimate (and also the minimum mean squared error estimate) \cite{anderson1979}. 
Suppose the channel is estimated with the help of \redtext{pilot symbols equally spaced in every $\nu$ samples}. The receive pilot signal vector can then be written as $\bv_{\floor{k/\nu}}$, where $\floor{\cdot}$ denotes the floor function and $\ba_k$ denotes a vector of length $k$.
\redtext{Applying the Kalman filter based on this pilot vector $\bv_{\floor{k/\nu}}$, the variance of the channel estimation error at the $\ell$-th pilot $\psi_\ell$ is given by the following recursive relations \cite{Medard2000}
\begin{align}
\label{eq:kalman_error1}
\frac{1}{\psi_1} &= \frac{1}{\sigma_h^2+\sigma_\xi^2} + \frac{\sigma_v^2}{\sigma_n^2} \\
\label{eq:kalman_error2}
\frac{1}{\psi_{\ell+1}} &= \frac{1}{\alpha^2\psi_\ell+\sigma_\xi^2} + \frac{\sigma_v^2}{\sigma_n^2},
\end{align}
where $\sigma_h^2$ is the channel variance, $\sigma_v^2$ is the pilot signal power assumed to be the same for all pilot symbols, and $\sigma_n^2$ is the noise power. \redtext{To explicitly express the channel estimation error, the channel is decomposed as
\begin{align}
h[i] = \bar{h}[i] + \tilde{h}[i],
\end{align}
where $\bar{h}[i]$ is the known part and $\tilde{h}[i]$ is the estimation error. The variance of the known part $\bar{h}[i]$ can be written as
\begin{align}
\sigma_{\bar{h}}^2[i] =  \sigma_h^2 - \sigma_{\tilde{h}}^2[i]. 
\end{align}
This notation is used in the derivation of the lower bound below.}
Note that the estimation error variances given in \eqref{eq:kalman_error1} and \eqref{eq:kalman_error2} are at the sampling points corresponding to the pilots. When they are used to decode the data part, the channel time variation will further degrade the estimation accuracy. This increase in estimation error is determined from the channel correlation function and the total estimation error variance at a given sampling point can be written as
\begin{align}
\sigma_{\tilde{h}}^2[i] = \psi_{\floor{i/\nu}}+(1-|R_h((i-\floor{i/\nu}\nu)T)|^2).
\end{align} }For very long sequence of signal, e.g., when $k\to \infty$, the error variance from the Kalman filter converges to some value $\psi$ (i.e., does not depend on the pilot index) given by \cite{Medard2000}
\begin{align}
\label{eq:k_infty_error}
\psi = & \frac{|R_h(\nu T)|^2-1-\SNR_v G_\rma(\theta)\sigma_\xi^2}{2\SNR_v G_\rma(\theta)|R_h(\nu T)|^2} \nonumber
\\ & + \frac{\sqrt{(|R_h(\nu T)|^2-1-\SNR_v G_\rma(\theta)\sigma_\xi^2)^2+4\SNR_v^2 G_\rma(\theta)\sigma_\xi^2 |R_h(\nu T)|^2}}{2\SNR_v G_\rma(\theta)|R_h(\nu T)|^2}.
\end{align} 
where $\SNR_v=\sigma_v^2/\sigma_n^2$ is the SNR of the pilot symbol excluding the antenna gain\redtext{. }$G_\rma(\theta)$ is the antenna gain compared to omnidirectional antenna \redtext{and is given by} 
\begin{align}
G_\rma(\theta) = \frac{G(\mu_\rmr|\mu_\rmr)}{1/(2\pi)}=  \frac{e^{1/\theta^2}}{I_0(1/\theta^2)}
\end{align}
where $1/(2\pi)$ in the denominator is the gain of the omnidirectional antenna, and  $G(\mu_\rmr|\mu_\rmr)$ is the peak of the antenna pattern with the main beam pointing at $\mu_\rmr$. $G(\alpha|\mu_\rmr)$ is assumed to have the shape of the von Mises PDF. Note that we use the peak of the antenna pattern here because the time scale of a packet is small and there will be negligible change in the pointing direction \redtext{within one packet}. 

Now consider the mutual information for only the $i$-th sample with channel estimate with error given in \eqref{eq:k_infty_error}. 
The worst case that the error $\tilde{h}[i]$ can have is to act as AWGN \cite{Medard2000}. In that case, the mutual information can be lower bounded by
\begin{align}
\label{eq:mi_single_lower}
I(s[i];y[i]|\bv_{\floor{i/\nu}}) \ge \ln \pa*{ 1 + \frac{\sigma_{\bar{h}}^2[i]\sigma_s^2}{\sigma_{\tilde{h}}^2\sigma_s^2+\sigma_n^2} }.
\end{align}
Using \eqref{eq:mi_single_lower}, and assuming the estimator does not use the decoded data for channel estimation and only use the pilot $\bv_{\floor{i/\nu}}$ then it can be shown that \cite{Medard2000}
\begin{align}
I(\bs_k;(\by_k, \bv_{\floor{k/\nu}} ) ) \ge \sum_{i\le k} I(s[i];y[i]|\bv_{\floor{i/\nu}} ).
\end{align} 
Plugging in the result so far, a lower bound for the mutual information can be written as
\begin{align}
I(\bs_k;(\by_k,\bv_{\floor{k/\nu}})) \ge  \sum_{i\le k}^{} \ln \pa*{ 1 + \frac{ (|R_h((i-\floor{i/\nu}\nu)T)|^2 - \psi_{\floor{i/\nu}}) \SNR_\rms G_\rma(\theta)}{(\psi_{\floor{i/\nu}} + ( 1 - |R_h((i-\floor{i/\nu}\nu)T)|^2)  )\SNR_\rms G_\rma(\theta) + 1 } } 
\end{align}
where $\SNR_\rms = \sigma_\rms^2/\sigma_n^2$ is the SNR of the data part excluding the antenna gain. 
Further assume $k\to \infty$ then $\psi_{\floor{i/k}}\to \psi$, and we have
\begin{align}
\label{eq:mi_lower_bound}
\lim\limits_{k\to \infty} \frac{1}{k} I(\bs_k;(\by_k,\bv_{\floor{k/\nu}})) & \ge \frac{1}{\nu} \sum_{i=2}^{\nu} \ln \pa*{ 1 + \frac{(|R_h(iT)|^2 - \psi) \SNR_\rms G_\rma(\theta)}{(\psi + ( 1 - |R_h(iT)|^2)  )\SNR_s G_\rma(\theta) + 1 } } \\
\label{eq:mi_lower_bound_def}
& = I_\mathrm{low}(\theta, \SNR_\rms,\nu) .
\end{align}
At high SNR or when beamwidth $\theta$ is small (i.e., the antenna gain $G_\rma(\theta)$ is large), then 
\begin{align}
I_\mathrm{low} (\theta, \SNR_\rms, \nu) 
\simeq \frac{1}{\nu} \sum_{i=2}^{\nu} \ln \pa*{ 1 + \frac{|R_h(iT)|^2 - \psi }{\psi + ( 1 - |R_h(iT)|^2)   } }
\end{align}
which implies that the loss due to the channel time-variation acts in the same way as the interference and it cannot be mitigated by increasing the transmit power. 

\begin{figure} 
\centering
\subfigure[$I_\mathrm{low}$ for different coherence bandwidths]{%
\includegraphics[width=0.47\columnwidth]{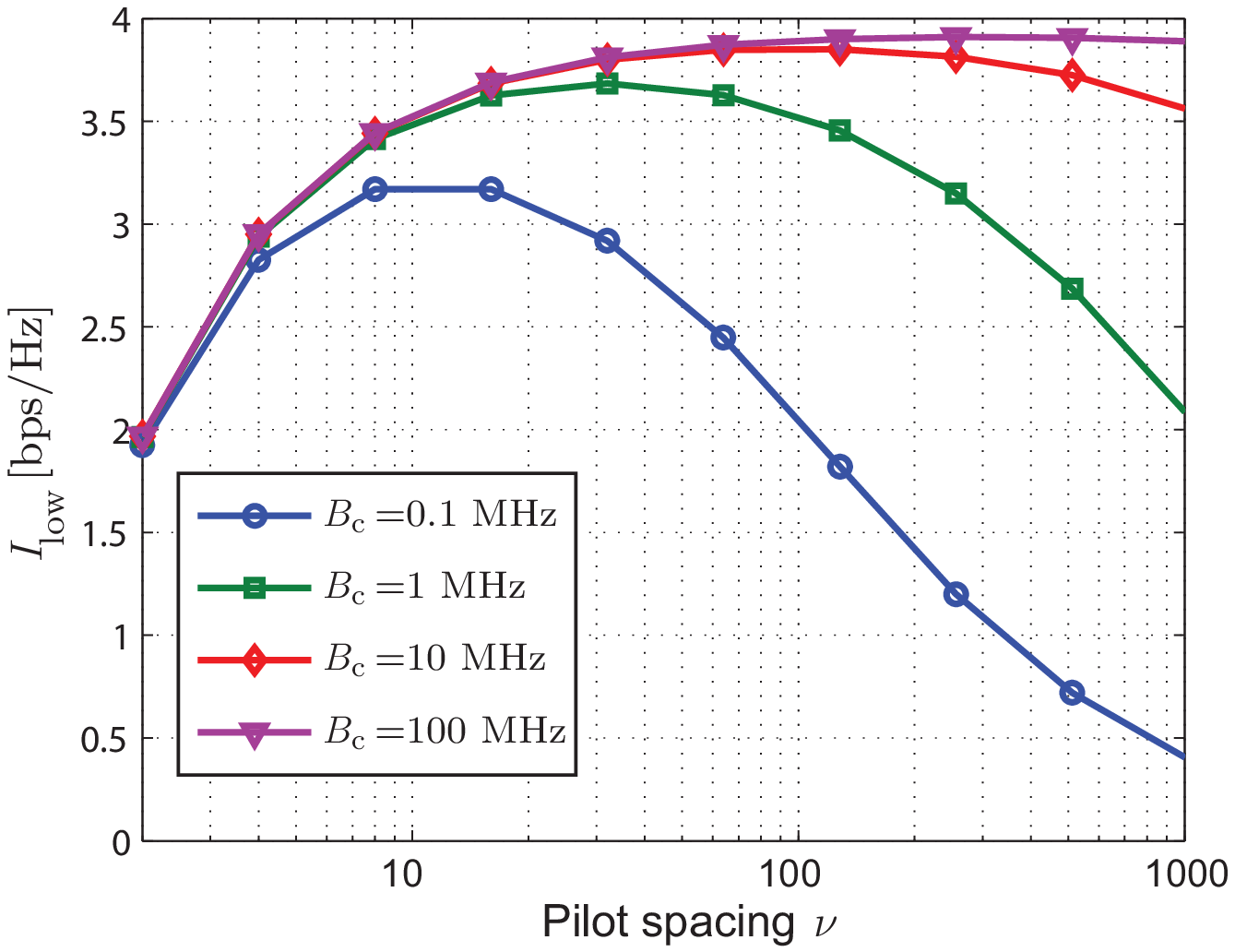}
\label{fig:lower_bound_func_nu_diff_BC}
}
\quad
\subfigure[$I_\mathrm{low}$ for different $\mu_\rmr$]{%
\includegraphics[width=0.47\columnwidth]{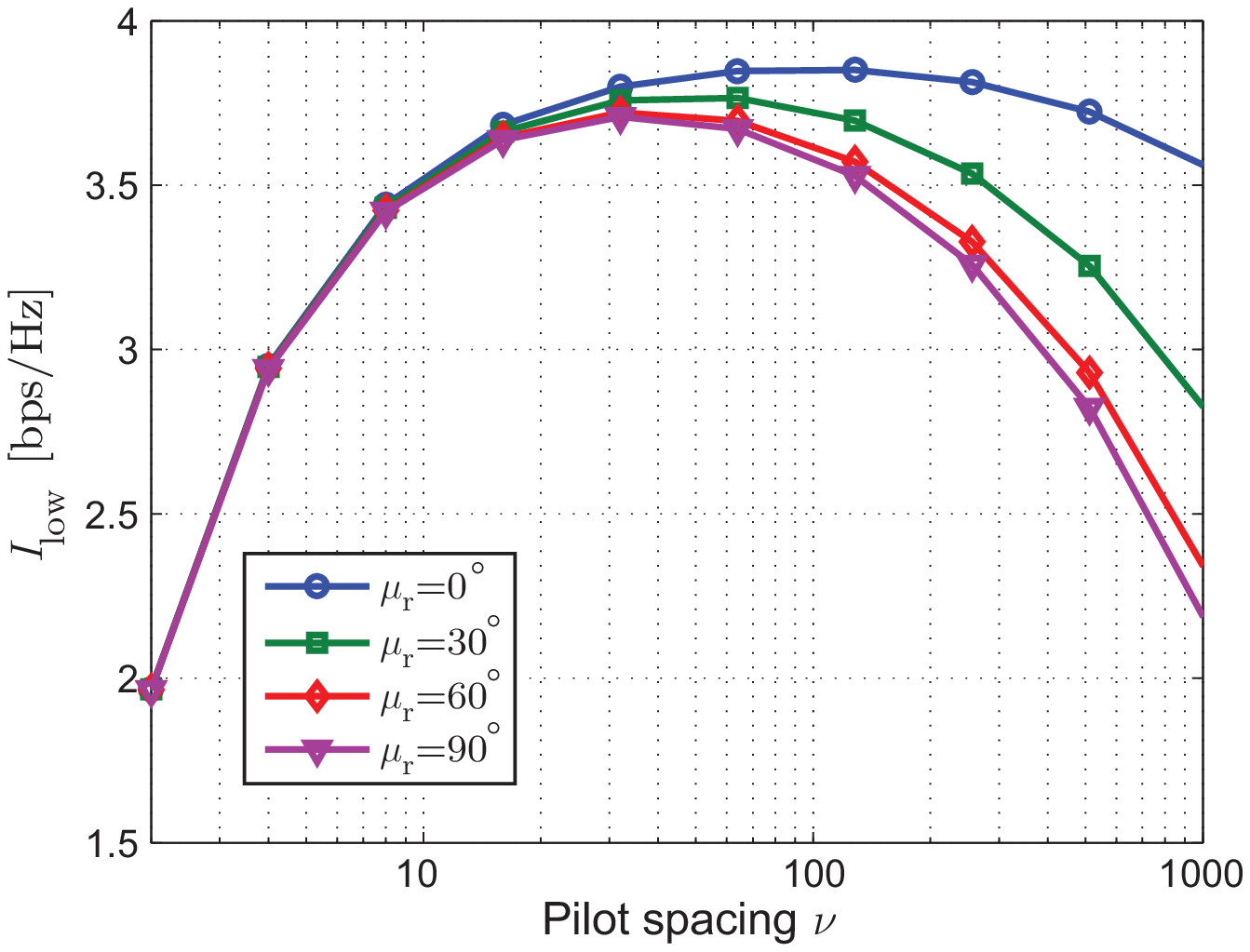}
\label{fig:lower_bound_func_nu_diff_mu_r}
}
\caption{Fig. \ref{fig:lower_bound_func_nu_diff_BC} shows the lower bound on the mutual information for a fixed beamwidth $\theta=10^\circ$ and different coherence bandwidth values (thus different $T$). In this example, $\SNR_\rms=\SNR_v=0$ dB, and $\mu_\rmr=0^\circ$ are used. In all cases, there exists an optimal value for the pilot spacing $\nu$. Fig. \ref{fig:lower_bound_func_nu_diff_mu_r} shows the lower bound on the mutual information for the same beamwidth $\theta=10^\circ$ and different angles of arrival $\mu_\rmr$. The coherence bandwidth is set to 10 MHz.}
\label{fig:mi_lower_bound}
\end{figure}

Fig. \ref{fig:lower_bound_func_nu_diff_BC} shows plots of the lower bound on the mutual information for different values of the coherence bandwidth against the pilot spacing $\nu$ (Note that $\nu$ determines the overhead for channel estimation. The larger $\nu$ is the lower the overhead.) In this example, $\SNR_\rms=\SNR_v=0$ dB, beamwidth is fixed at $\theta=10^\circ$, and $\mu_\rmr=0^\circ$ are used. We can see that in all these cases, there is an optimal $\nu$ that maximizes the lower bound. As the coherence bandwidth increases (thus $T=1/B_\mathrm{c}$ decreases), the time variation measured in the number of samples increases and the optimal $\nu$ increases.  

Fig. \ref{fig:lower_bound_func_nu_diff_mu_r} shows plots of the lower bound on the mutual information for different angles of arrival (in reference to the direction of travel) for a fixed beamwidth $\theta=10^\circ$, and the coherence bandwidth is set to $B_\mathrm{c}=10$ MHz. Remember that the correlation function depends on $\mu_\rmr$ and the correlation decreases slower as $|\mu_\rmr|$ becomes smaller as shown in Fig. \ref{fig:corr_kr=50_Dr=100}. Again, we can see there exists an optimal $\nu$ and this optimal $\nu$ increases as the angle of arrival $\mu_\rmr$ decreases. 


\subsection{How Often Should the Beams Be Realigned?} \label{sec:realignment_duration}
In this subsection, 
we want to find out the most appropriate time duration between beam realignments. We consider the beam sweeping as a method to align the beams. Two possible choices for the time duration between realignments are the channel coherence time $\Tc$ (Section \ref{sec:chan_coherence_time}) and the beam coherence time $\TB$ (Section \ref{sec:beam_coherence_time}). Assuming no error in the beam measurement during the alignment process, realignment in every $\Tc$ will ensure that the best beams, which provide the highest receive power, are always chosen. If realignment is done in every $\TB$ instead, suboptimal beams could result due to the effect of fading. The overhead is of course higher when realigning in every $\Tc$ than when realigning in every $\TB$ because $\TB\ge \Tc$ . We call the realignment in every $\Tc$ the short-term realignment and the realignment in every $\TB$ the long-term realignment. In the following, we will investigate the performance of these two cases. For the LOS channel, $\Tc$ and $\TB$ are of comparable values (see \eqref{eq:Tc_los} and \eqref{eq:TB_los}), and there is not much difference between the two. Therefore we study the NLOS case only in the followings. 

For clarity, 
we consider a channel with only two paths.
Denote $\Delta\ge 1$ the path loss ratio between the first and second path, $\PL_{i}$ for $i=\{1,2\}$ the path losses of the two paths, then
\begin{align}
\PL_{1} & =  \PL_{2} \, \Delta ,
\end{align}
where we have assumed without loss of generality that the first path has higher average receive power.
Let $g_{i}=|h_i|^2$ and $P_{i}$ where $i=\{1,2\}$ be the fading and the instantaneous receive power, respectively, then we have
\begin{align}
P_{i} = g_{i} \PL_{i}.
\end{align} 
Note that our channel model in \eqref{eq:channel_model_NLOS} corresponds to the fading coefficient and no path loss was incorporated. 

The beam sweeping will select a beam following the rule $i^\star = \arg\max_{i} P_{i}$, explicitly
\begin{align}
i^\star 
& = \begin{cases}
1 & \text{if } g_{1} \ge g_{2}/\Delta \\
2 & \text{if } g_{1}<g_{2}/\Delta
\end{cases}.
\end{align}
Let $f_g(g)$ be the PDF of $g_{i}$, then the beam sweeping will output 1 and 2 with probabilities
\begin{align}
\bbP\{i^\star=1\} &= \int_{0}^{\infty} \int_{g_2/\Delta}^{\infty} f_g(g_1) \rmd g_1 f_g(g_2) \rmd g_2, \\
\bbP\{i^\star=2\} &= \int_{0}^{\infty} \int_{0}^{g_2/\Delta} f_g(g_1) \rmd g_1 f_g(g_2) \rmd g_2,
\end{align}
respectively. 

To have tractable analysis, we assume the fading is Rayleigh so that $g_i$ follows an exponential distribution with unit mean. When realigning in every $\Tc$, the path yielding the highest power is always chosen, so that the receive power follows the distribution of $\max\{P_1,P_2\}$. The SNR is proportional to the received power, and the SNR PDF can be derived as 
\begin{align}
f_\mathrm{short}(\gamma) = \frac{1}{\bar{\gamma}_1}e^{-\gamma/\bar{\gamma}_1} + \frac{1}{\bar{\gamma}_2}e^{-\gamma/\bar{\gamma}_2} - \left(\frac{1}{\bar{\gamma}_1} +\frac{1}{\bar{\gamma}_2}\right)e^{-(1/\bar{\gamma}_1 + 1/\bar{\gamma}_2)\gamma},
\end{align}
where $\bar{\gamma}_i=\frac{P_i}{P_\mathrm{n}}$ is the average SNR of the $i$-th path and $P_\mathrm{n}$ is the noise power. 
When realigning in every $\TB$, the beam sweeping is performed at the beginning and the selected beam will be used until the next realignment. Note that $\TB \gg \Tc$ for the NLOS channels (see numerical examples in Section \ref{sec:chan_coherence_time} and \ref{sec:beam_coherence_time}). The fading coefficient becomes uncorrelated after $\Tc$, thus the beam selected at the beginning could result in suboptimal receive power. Depending on the result of the beam sweeping the channel experienced here follows either $P_1$ or $P_2$. The SNR in this case follows
\begin{align}
f_\mathrm{long}(\gamma) = \frac{1}{\bar{\gamma}_i}e^{-1/\bar{\gamma}_i}.
\end{align}
So far we have derived the distribution of the SNR for the short- and long-term realignment. Now we will discuss the overhead of the two realignment durations. The time needed for beam sweeping is the same for both the short- and long-term realignments. Denoting this time duration by $T_\mathrm{sw}$, then the temporal efficiencies of the short- and long-term realignments are
\begin{align}
\eta_\mathrm{short}(\theta) &= \frac{\Tc(\theta) - T_\mathrm{sw}(\theta)}{\Tc(\theta)} \\
\eta_\mathrm{long}(\theta) &= \frac{\TB(\theta) - T_\mathrm{sw}(\theta)}{\TB(\theta)}.
\end{align}
Note that all these are functions of the beamwidth $\theta$. 

Finally, the loss due to the channel time-variation, the temporal efficiency, and the bound on the mutual information are all considered for the overall performance metric, i.e.,   
\begin{align}
\label{eq:spec_eff_short}
C_\mathrm{short}(\theta) & = \eta_\mathrm{short}(\theta) 
\bbE_{\mathrm{short}} \left[ I_\mathrm{low}(\theta,\gamma_\mathrm{short},\nu) \right] \\
\label{eq:spec_eff_long}
C_\mathrm{long}(\theta) & = \eta_\mathrm{long}(\theta) 
\bbE_{\mathrm{long}}\left[ I_\mathrm{low}(\theta,\gamma_\mathrm{long},\nu) \right] \\
 & = \eta_\mathrm{long}(\theta) \left( \bbP\{i^\star=1\}\bbE_{\gamma_1} \left[ I_\mathrm{low}(\theta,\gamma_1,\nu) \right] + \bbP\{i^\star=2\}\bbE_{\gamma_2} \left[ I_\mathrm{low}(\theta,\gamma_2,\nu) \right] \right)
\end{align}
where  
$I_\mathrm{low}(\theta,\gamma,\nu)$ is the lower bound derived in \eqref{eq:mi_lower_bound_def} in the previous subsection. 

\begin{figure}
\centering
\includegraphics[width=0.6\textwidth]{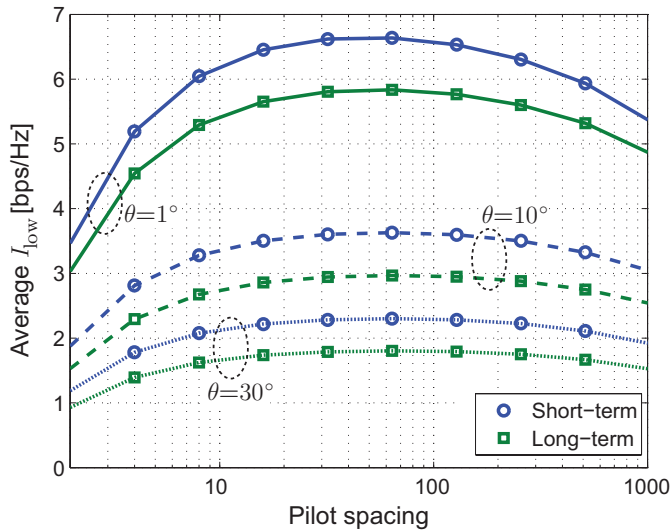}
\caption{Average $I_\mathrm{low}$ for different beamwidths for both the short- and long-term realignment. The sensitivity to the pilot spacing near the optimal point is low and there is little dependence on the beamwidth. For $\theta$ up to $30^\circ$, pilot spacing of $64$ is optimal or has negligible loss.}
\label{fig:Average_I_low_diff_beamwidth}
\end{figure}

\begin{figure} 
\centering
\subfigure[Path loss ratio $\Delta=3$ dB]{%
\includegraphics[width=0.47\columnwidth]{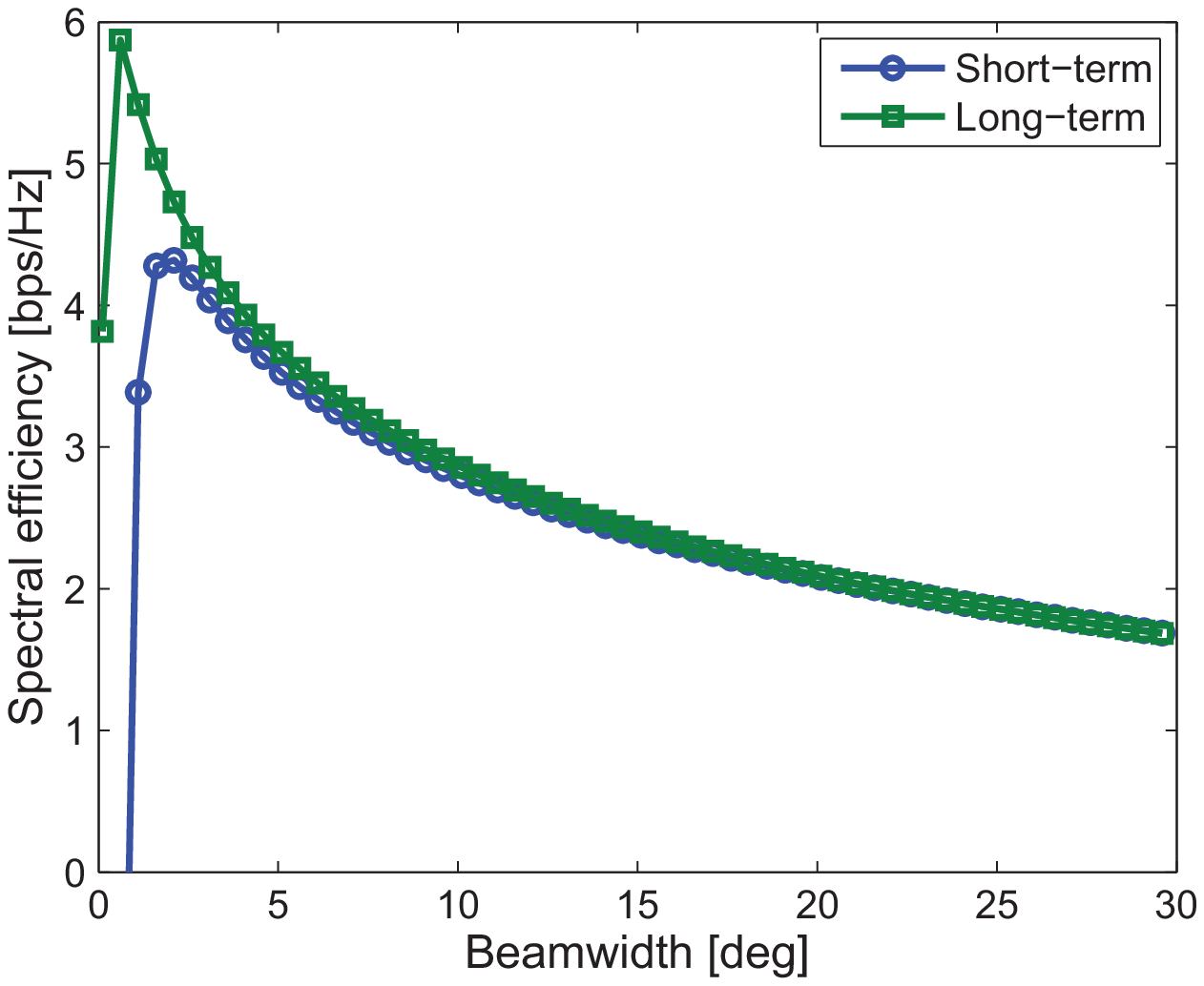}
\label{fig:spectral_eff_3c_delta=3dB}
}
\quad
\subfigure[Path loss ratio $\Delta=10$ dB]{%
\includegraphics[width=0.47\columnwidth]{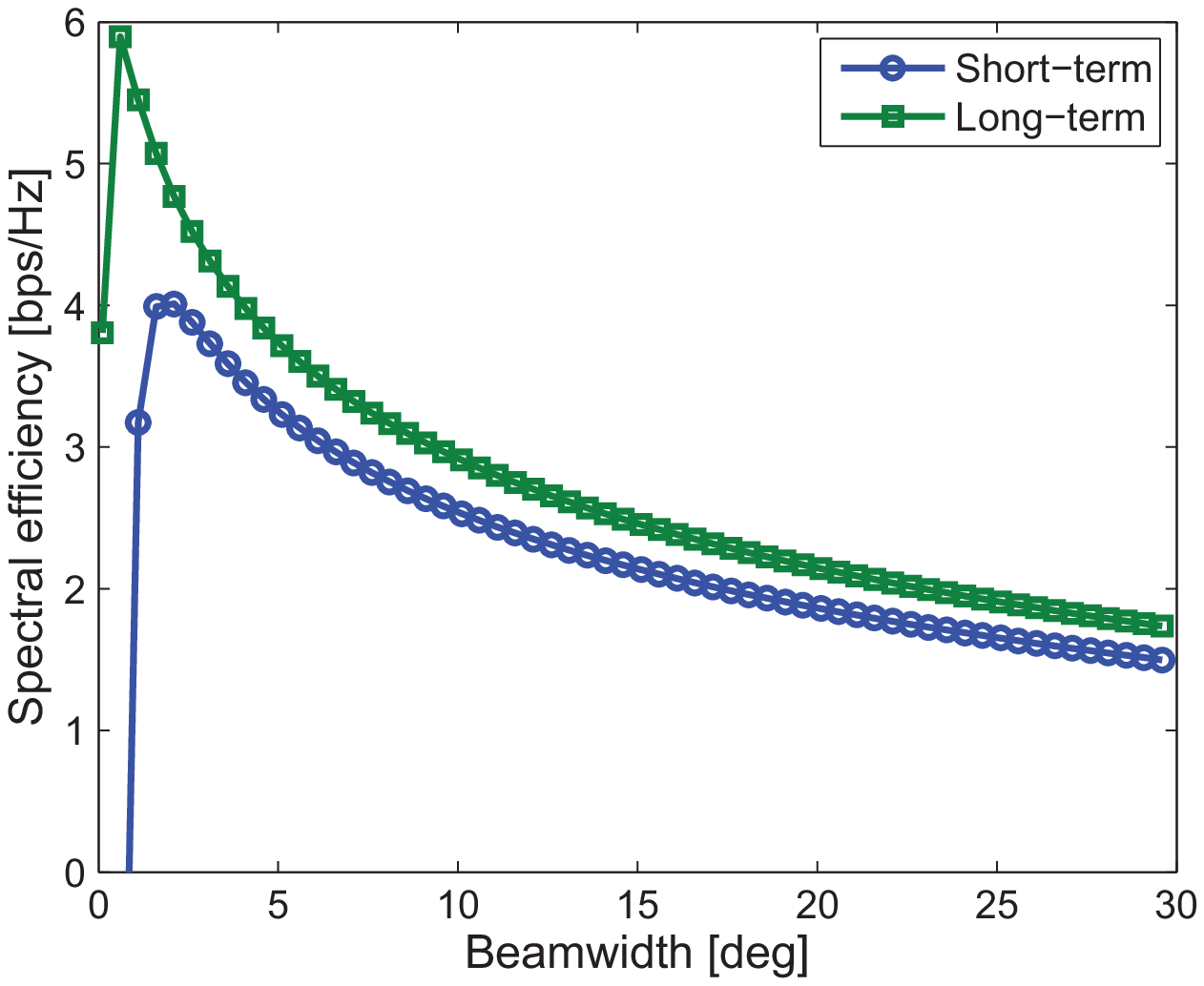}
\label{fig:spectral_eff_3c_delta=10dB}
}
\caption{Comparison of the spectral efficiencies in \eqref{eq:spec_eff_short} and \eqref{eq:spec_eff_long} for the short- and long-term beam realignment when the beam sweeping follows the 802.15.3c method. Fig. (a) and (b) show the case when the path loss ratio of the two path $\Delta$ is $3$ dB and $10$ dB, respectively. In both cases the long-term realignment performs better and the gap is more pronounced when $\Delta=10$ dB. The gap increases for larger $\Delta$ because the sweeping is less likely to make mistake when $\Delta$ is large so that the large overhead of the short-term realignment penalizes rather than improves the performance.}
\label{fig:spectral_eff_3c}
\end{figure}

Now we provide a numerical example comparing the spectral efficiencies in \eqref{eq:spec_eff_short} and \eqref{eq:spec_eff_long} when realignment duration is set to $\Tc$ versus $\TB$. To make the comparison meaningful, the pilot spacing $\nu$ should be optimized for all $\theta$. Fig. \ref{fig:Average_I_low_diff_beamwidth} shows plots for different $\theta$ for both the short- and long-term realignment. As can be observed from the plots, the sensitivity to $\nu$ is small near the optimal $\nu$, and in this case $\nu=64$ is optimal or has negligible loss for all the range of beamwidth up to $30^\circ$. Based on this we set $\nu=64$, and compute the spectral efficiencies as a function of $\theta$. For the beam sweeping, we consider a basic approach adopted in the IEEE 802.15.3c \cite{Wang2009} which is based on hierarchical beam codebook. Let $\ell$ the number of levels in the codebook, and the $i$-level has $L_i$ beams. In this approach, at each level all the beam combination pairs are tested, so the overhead of beam training is $L_i^2T_\mathrm{TRN}$ for the search at the $i$-th level. $T_\mathrm{TRN}$ is the duration needed for one beam measurement. Thus the 802.15.3c method has overhead of $T_\mathrm{3c}=T_\mathrm{TRN}\sum_{i=1}^{1/\ell}L_i^2$. It can be shown that the optimum $L_i$ that minimize the number of beam training is when $L_1=\dots=L_\ell=L=\left(\frac{\theta_0}{\theta} \right)^{1/\ell}$, where $\theta_0$ is the coverage and $\theta$ is the desired beamwidth. In this case, the overhead can be expressed as
\begin{align}
T_\mathrm{3c}(\theta) = \ell \left( \frac{\theta_0}{\theta} \right)^{2/\ell}T_\mathrm{TRN}.
\end{align}
Note that the overhead here ignores the acknowledgment phase. 
Plugging in $T_\mathrm{SW}(\theta)=T_\mathrm{3c}(\theta)$, we can now compute the spectral efficiencies in \eqref{eq:spec_eff_short} and \eqref{eq:spec_eff_long} as a function of the beamwidth $\theta$. The coherence bandwidth is set to 10 MHz, pointing angle $\mu_\rmr=90^\circ$ (which corresponds to the worst case), $\theta_0=180^\circ$, the training per beam $T_\mathrm{TRN}=1\mu$s, and angular spread $\sigma_\mathrm{AS}=25.7^\circ$. Other parameters are the same as used in the previous subsection. The result is shown in Fig. \ref{fig:spectral_eff_3c} for the case when the path loss ratio $\Delta$ is $3$ dB and $10$ dB. In both cases, the long-term realignment has higher spectral efficiency and the gap is larger for large $\Delta$. This is because when $\Delta$ is large, the probability that beam sweeping chooses the suboptimal choice becomes smaller so that minimal benefit can be expected from the short-term realignment. Thus the overhead paid for the short-term realignment does not provide sufficient return and the long-term realignment performs better due to the lower required overhead.  

\section{Conclusion} \label{sec:conclusion}
In this paper, we derived the channel coherence time as a function of the beamwidth taking both Doppler effect and pointing error into consideration. Our results  show that there exists a non-zero optimal beamwidth that maximizes the channel coherence time. If the beamwidth is too narrow, pointing error will limit the coherence time. If the beamwidth is too wide, the Doppler spread becomes the limiting factor. 

We defined and computed a new quantity called the beam coherence time, which is tailored to the beam alignment context. We showed that the beam coherence time is typically an order-of-magnitude longer than the conventional coherence time. To reduce the impact of the overhead of doing realignment in every channel coherence time, we showed that beams should be realigned every beam coherence tiime for the best performance. 

A natural next step in our work is to compute beam coherence times from measured channel data. To the best of our knowledge, existing mmWave measurements in the vehicular context have focused mostly on characterization of the pathloss as the measurement apparatus have limited control over the beamwidth used \cite{Thomas1994,Takahashi2003,Yamamoto2008}. It would be useful to develop and categorize beam coherence times for different environments. 


%



%
%
%
%



\bibliographystyle{IEEEtran}
\bibliography{ref}

%

%
%
%




\end{document}

%% file: new_command.tex
\usepackage{mathtools}

\newcommand*{\ba}{\boldsymbol{a}}

\newcommand*{\bs}{\boldsymbol{s}}

\newcommand*{\bv}{\boldsymbol{v}}

\newcommand*{\by}{\boldsymbol{y}}

\newcommand*{\rma}{\mathrm{a}}

\newcommand*{\rmc}{\mathrm{c}}
\newcommand*{\rmd}{\mathrm{d}}

\newcommand*{\rmr}{\mathrm{r}}

\newcommand*{\jj}{\mathrm{j}}

\newcommand*{\SNR}{{\mathsf{SNR}}}

\newcommand*{\bbE}{\mathbb{E}} 
\newcommand*{\bbP}{\mathbb{P}} 

\newcommand*{\mcP}{\mathcal{P}}

\newcommand*{\Gauss}{\mathcal{N}} 

\DeclarePairedDelimiter{\floor}{\lfloor}{\rfloor}
\DeclarePairedDelimiter{\curlbr}{\lbrace}{\rbrace}
\DeclarePairedDelimiter{\sqbr}{\lbrack}{\rbrack}
\DeclarePairedDelimiter{\pa}{\lparen}{\rparen}
